\documentclass{WileyMSP-template}
\usepackage{xcolor}
\usepackage{amsmath}
\usepackage{upgreek}
\usepackage{ulem}

\begin{document}

\pagestyle{fancy}
\rhead{\includegraphics[width=2.5cm]{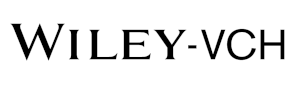}}

\title{Short-Term Bienenstock-Cooper-Munro Learning in Optoelectrically-Driven Flexible Halide Perovskite \\ Single Crystal Memristors}

\maketitle

\author{Ivan Matchenya, $^{1,2}$}
\author{Anton Khanas, $^{3}$}
\author{Roman Podgornyi, $^{1}$}
\author{Daniil Shirkin, $^{1}$}
\author{Alexey Ekgardt, $^{1}$}
\author{Nikita Sizykh, $^{3}$}
\author{Sergey Anoshkin, $^{1}$}
\author{Dmitry V. Krasnikov, $^{2}$}
\author{Alexei Yulin, $^{1}$}
\author{Alexey Zhukov, $^{4}$}
\author{Albert G. Nasibulin, $^{2}$}
\author{Ivan Scheblykin, $^{5}$}
\author{Anatoly Pushkarev, $^{2,*}$}
\author{Andrei Zenkevich, $^{3,*}$}
\author{Juan Bisquert, $^{6,*}$}
\author{Alexandr Marunchenko, $^{1,5,*}$}

\begin{affiliations}
$^1$ ITMO University, School of Physics and Engineering, St. Petersburg, 197101, Russian Federation

$^2$ Skolkovo Institute of Science and Technology, 30/1 Bolshoy Boulevard, 121205 Moscow, Russian Federation

$^3$ Moscow Institute of Physics and Technology (National research university), Institutskiy per.
9, Dolgoprudny, Moscow Region, 141701, Russia

$^4$ International Laboratory of Quantum Optoelectronics, HSE University, 190008, St. Petersburg, Russian Federation

$^5$ Chemical Physics and NanoLund, Lund University, P.O. Box 124, 22100 Lund, Sweden

$^6$ Instituto de Tecnolog\'ia Qu\'imica (Universitat Polit\`ecnica de Val\`encia-Agencia Estatal Consejo Superior de Investigaciones Cient\'ificas), 46022 Val\`encia, Spain

Andrey Zenkevich\\
Email Address:
zenkevich.av@mipt.ru

A.P. Pushkarev\\
Email Address:
anatoly.pushkarev@metalab.ifmo.ru

Juan Bisquert\\
Email Address:
bisquert@uji.es\\

A.A. Marunchenko\\
Email Address:
a.marunchenko@metalab.ifmo.ru\\

\end{affiliations}


\begin{abstract}

\par The transition to smart, wearable, and flexible optoelectronic devices that communicate with each other and perform neuromorphic computing at the edge is a major goal in next-generation optoelectronics. These devices are expected to carry out their regular tasks while being supported by energy-efficient, in-memory computations. In this study, we present a lateral flexible device based on cesium lead tribromide metal halide perovskite single crystals, integrated with single-walled carbon nanotube thin-film electrodes. We demonstrate that our device follows the Bienenstock-Cooper-Munro theory of synaptic modification under hybrid optoelectronic stimuli. This biorealistic response paves the way for the development of hybrid organic-inorganic artificial visual systems.
\end{abstract}

\section{Introduction}

\par Modern computers based on the classical Von Neumann architecture are reaching energy consumption limits in supporting artificial intelligence (AI) algorithms and Internet of Things (IoT) applications  \cite{christensen2022, conklin2023solving}. Neuromorphic computing has emerged as an alternative paradigm inspired by the structure and operation of the biological brain \cite{markovic2020physics, song2023recent, ielmini2021brain}. The complexity of neurons and their connections and the high level of parallelism are the reasons for the high energy efficiency of biological organisms compared to classical computers\cite{lynn2019physics, chaurasiya2023emerging, kumar2022dynamical, kumar2020third}. Within this framework, memristors are functional resistive elements capable of simultaneously mimicking the behavior of both neurons and synapses. As a result, they are considered fundamental building blocks for next-generation neuromorphic computing hardware \cite{kumar2022dynamical, chua1976memristive, chaurasiya2023emerging, bisquert2023device, mikheev2019ferroelectric, khanas2022second}.

\begin{figure*}[t]
	\centering
	\includegraphics[scale = 0.85]{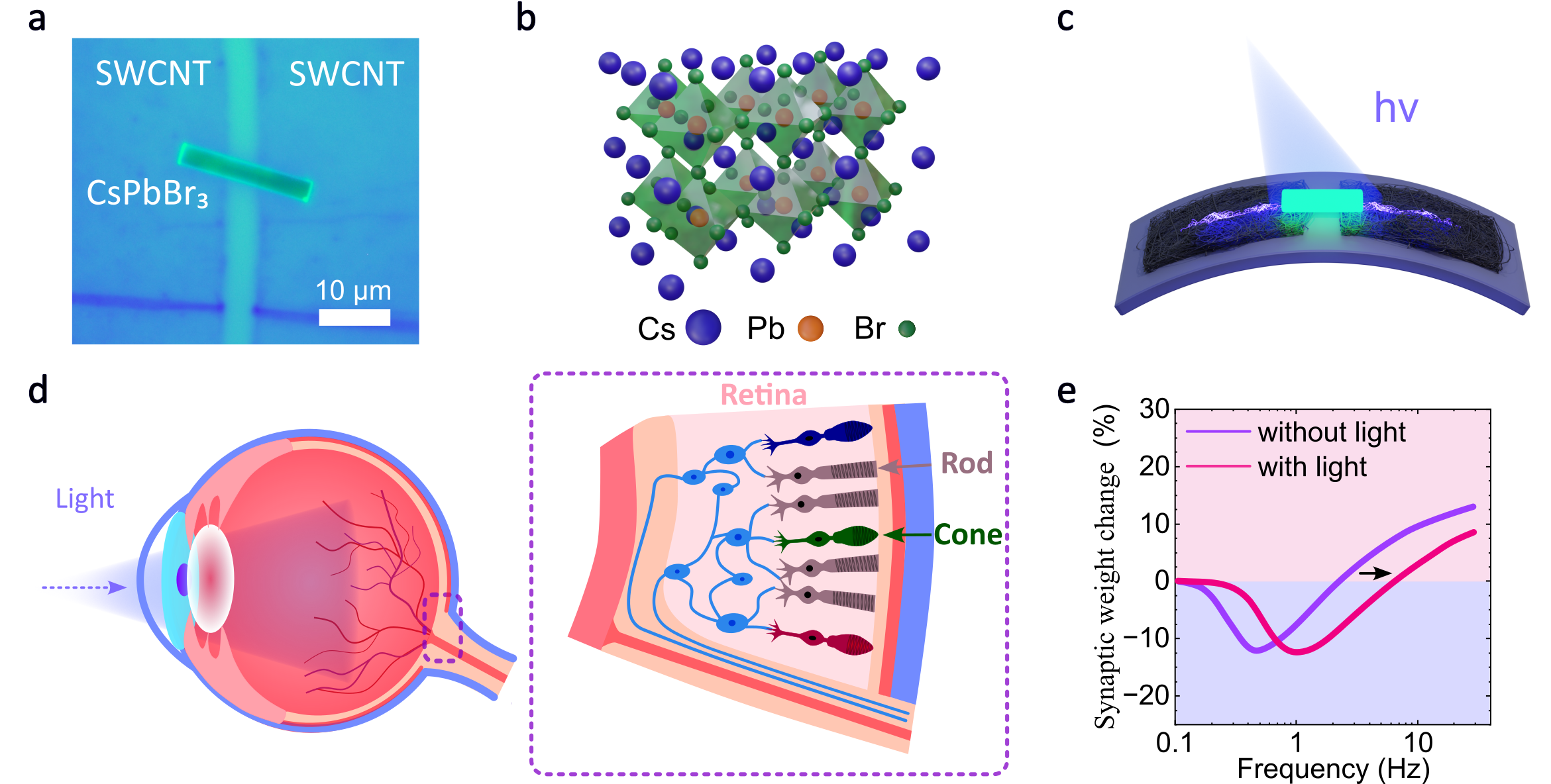}
	\caption{\textbf{Visual system based on CsPbBr$_3$ single-crystal following the Bienenstock-Cooper-Munro rule.} \textbf{(a)} Microscopy image of a perovskite single-crystal on the flexible substrate connected to symmetrical SWCNT  electrodes (photoluminescent image of the actual device).
		\textbf{(b)} The orthorhombic crystal structure of the employed CsPbBr$_3$ microcrystal. \textbf{(c)} Operation of the single-crystal device under combined optoelectrical stimuli and mechanical bending. \textbf{(d)} Schematic of the human visual system: light is transmitted to the retina, where light-sensitive cones and rods send a collective response signal to the brain. \textbf{(e)}  Biological response of the device under hybrid optoelectrical stimuli, following the BCM rule. The synaptic weight change is frequency-dependent, exhibiting a depression region at low frequencies, followed by facilitation at high frequencies of internal input pulses. The sliding threshold, which separates depression and facilitation, is influenced by light stimuli.}
	\label{fig:FIG.1}
\end{figure*}

\par The complex nonlinear synaptic functionality of memristive elements plays a crucial role in emulating biological neural networks. One notable example is the Bienenstock-Cooper-Munro (BCM) rule for synaptic modification, which is naturally realized in the visual cortex \cite{cooper2012bcm, rittenhouse1999monocular, frenkel2004monocular, coleman2010rapid}. Within the framework of BCM theory, it is shown that the response of the nervous tissue to a sequence of electrical pulses nonmonotonically depends on the pulse repetition frequency (Fig.\ref{fig:FIG.1}e). Initially, the response decreases at low frequencies, corresponding to synaptic depression. It then reaches a minimum point before increasing again as the frequency rises. Importantly, there exists a specific threshold frequency at which no synaptic modification occurs. This threshold is dynamically adjusted based on the characteristic intensity of light in the living organism’s environment — specifically, how actively these brain regions are engaged in processing visual information \cite{ bienenstock1982theory, cooper2004theory, cooper2012bcm}. 

\par Memristive devices implementing this learning mechanism require at least second-order dynamic behavior \cite{cooper2012bcm, john2022ionic} and can serve as functional components in neuromorphic hardware for image recognition tasks \cite{wang2020toward, john2022ionic, jiang2023mammalian}. \textcolor{black}{Moreover, incorporating the BCM learning rule in spiking neural networks (SNNs) has been shown to significantly reduce training time \cite{xu2016hardware}, enhance network sparsity (thus improving energy efficiency) \cite{xu2017improving}, and outperform conventional spike-timing-dependent plasticity (STDP) models in SNNs with high neuronal connectivity \cite{galluppi2015framework}. Ultimately, the BCM learning rule plays a pivotal role in advancing neuromorphic computing and enabling new approaches to real-time learning in artificial neural networks.}

\par It is important that the BCM rule be implemented as directly as it is in living organisms to simplify the realization of neural network algorithms. Current research demonstrates BCM functionality in two different ways (Table 1). The first approach involves the preliminary stimulation of a memristor using rate-based spike trains that emulate synaptic activity under different sensory conditions. However, in this case, the synaptic weight change remains monotonic across the entire frequency range \cite{du2015biorealistic, khanas2022second, xiong2019bienenstock} lacking the characteristic synaptic depression region  \cite{cooper2012bcm}. To address this limitation, a \textcolor{black}{triplet spike-timing-dependent plasticity (triplet-STDP)} scheme has been designed \cite{wang2020toward, john2022ionic}. Here, a sequence of three voltage pulses where one of the pulses differs in polarity from the other two, is received at the input of the device. Depending on the pulse repetition frequency (time intervals between spikes), the conductivity value of the device changes according to the BCM learning rule. Thus, a more direct BCM implementation that also incorporates a response to light stimuli, similar to that observed in the visual cortex of biological organisms, would be highly beneficial for further device development. This approach would allow computing tasks related to image recognition (involving a light source) to be processed naturally in-materia \cite{milano2022materia, zhang2022sensor}, leveraging the advantages of the optical computing paradigm \cite{mcmahon2023physics}.

\begin{table}[t]
    \centering
    \begin{tabular}{ccccc}
    \hline
       \text{Materials}  & \text{Operation} & \text{BCM method} & \text{Timescale} & Report \\
         \hline
        MAPbI$_3$ & Electrical & Triplet STDP & Long-term & \cite{john2022ionic}\\
        WO$_{3-x}$ & Electrical & Triplet STDP & Long-term & \cite{wang2020toward}\\
        STO, Nb-STO & Electrical & Priming pulses & Short-term & \cite{xiong2019bienenstock} \\
        CsFAPbI$_3$, CsPbBr$_2$I  & Optoelectrical & Triplet STDP & Long-term & \cite{ren2022synaptic} \\
         \textcolor{black}{$\alpha$-IGZO}  & \textcolor{black}{Electrical} & \textcolor{black}{Priming pulses} & \textcolor{black}{Short-term} & \textcolor{black}{\cite{zhu2023implementation}}\\
         \textcolor{black}{$\alpha$-IGZO}  & \textcolor{black}{Optoelectrical} & \textcolor{black}{Priming pulses} & \textcolor{black}{Short-term} & \textcolor{black}{\cite{ke2022bcm}}\\
         \textcolor{black}{ITO/GEL-CCs/ITO}  & \textcolor{black}{Electrical} & \textcolor{black}{Priming pulses} & \textcolor{black}{Short-term} & \textcolor{black}{\cite{chen2024proton}}\\
         \textcolor{black}{ITO/$\lambda$-car/ITO}  & \textcolor{black}{Electrical} & \textcolor{black}{Priming pulses} & \textcolor{black}{Short-term} & \textcolor{black}{\cite{guo2020bienenstock}}\\
        CsPbBr$_3$ single crystal & Optoelectrical & Direct & Short-term & This work\\
        \hline
    \end{tabular}
            \caption{Realization of BCM functionality in memristor devices.}
    \label{tab:bcm_types}
\end{table}
  
\par Metal halide perovskites are a decent class of semiconductor materials exhibiting intrinsically complex charge carrier dynamics. Previously, they have been comprehensively studied aiming at not only optoelectronic applications such as photodetectors and solar cells \cite{marunchenko2022single, noel2014lead, stranks2015metal}, but also for memory applications \cite{marunchenko2024memlumor, marunchenko2024charge, john2022reconfigurable, john2021halide, park2022metal} due to their dynamic nature of defects and mobile ions \cite{marunchenko2024memlumor, marunchenko2024charge, marunchenko2023mixed, john2022ionic, sakhatskyi2022assessing}. Noticeably, the signature of the 2nd-order dynamics combined with the BCM learning rule has been revealed in multilayer device based on \textcolor{black}{methylammonium lead idoide} (MAPbI$_3$) perovskite by John et al., where the triplet spike-timing-dependent plasticity scheme was realized \cite{john2022ionic} (Table 1). By integrating both memory functionality and light responsiveness, metal halide perovskites emerge as a promising material platform for the development of optoelectronic visual systems.

\par In this work, we demonstrate the short-term BCM response of cesium lead tribromide halide perovskite (CsPbBr$_3$) microcrystals to the hybrid optoelectrical stimuli. The photosensitive microcrystals are electrically connected to single-walled carbon nanotube (SWCNT) thin-film electrodes on a flexible substrate, mimicking the cones embedded in the flexible human retina (Fig.\ref{fig:FIG.1}). Our results establish that in these devices, both frequency-dependent potentiation and depression, observed in both dark conditions and under light illumination, follow the Bienenstock-Cooper-Munro (BCM) learning rule. Moreover, we demonstrate that this BCM functionality, particularly in the depression region, does not require specialized electrical pulses in the form of triplets or presynaptic rate-based pulse trains (Table 1). Such a behavior upon hybrid optoelectronic stimuli can be explained by the competition of the capacitive- and inductive-like contributions coexisting in the charge carrier dynamics of metal halide perovskites. 

\section{Results and Discussion}

\par Our study focuses on the lateral metal-semiconductor-metal type of the optoelectronic device \cite{marunchenko2022single, marunchenko2023mixed}. In such geometry, we can directly probe the dynamics of halide perovskite semiconductors as compared to the typical vertical device structure with additional charge transport layers. To enable such a structure, we use a chemically inert SWCNT thin film as an electrode material and CsPbBr$_3$ orthorhombic perovskite single-crystal as the semiconducting functional material (Fig.\ref{fig:FIG.1} a-c) similar to our previous works \cite{marunchenko2022single, marunchenko2023mixed}. When assembled on a flexible substrate, our crystals can be viewed as artificial cones embedded in the retinal layer of the human eye (Fig. \ref{fig:FIG.1} d). An additional advantage of halide perovskite single crystals over polycrystalline samples is their superior charge transport properties \cite{marunchenko2022single, zhumekenov2016formamidinium, li2018metal, feng2018single, xia2021limits}, as demonstrated by recent advancements in large-scale integration technologies for optoelectronic vision applications \cite{zhang2022large, yang2023towards, zhang2022vertical, gu2023planar}. 

\begin{figure*}[h!]
	\centering
	\includegraphics[scale = 0.75]{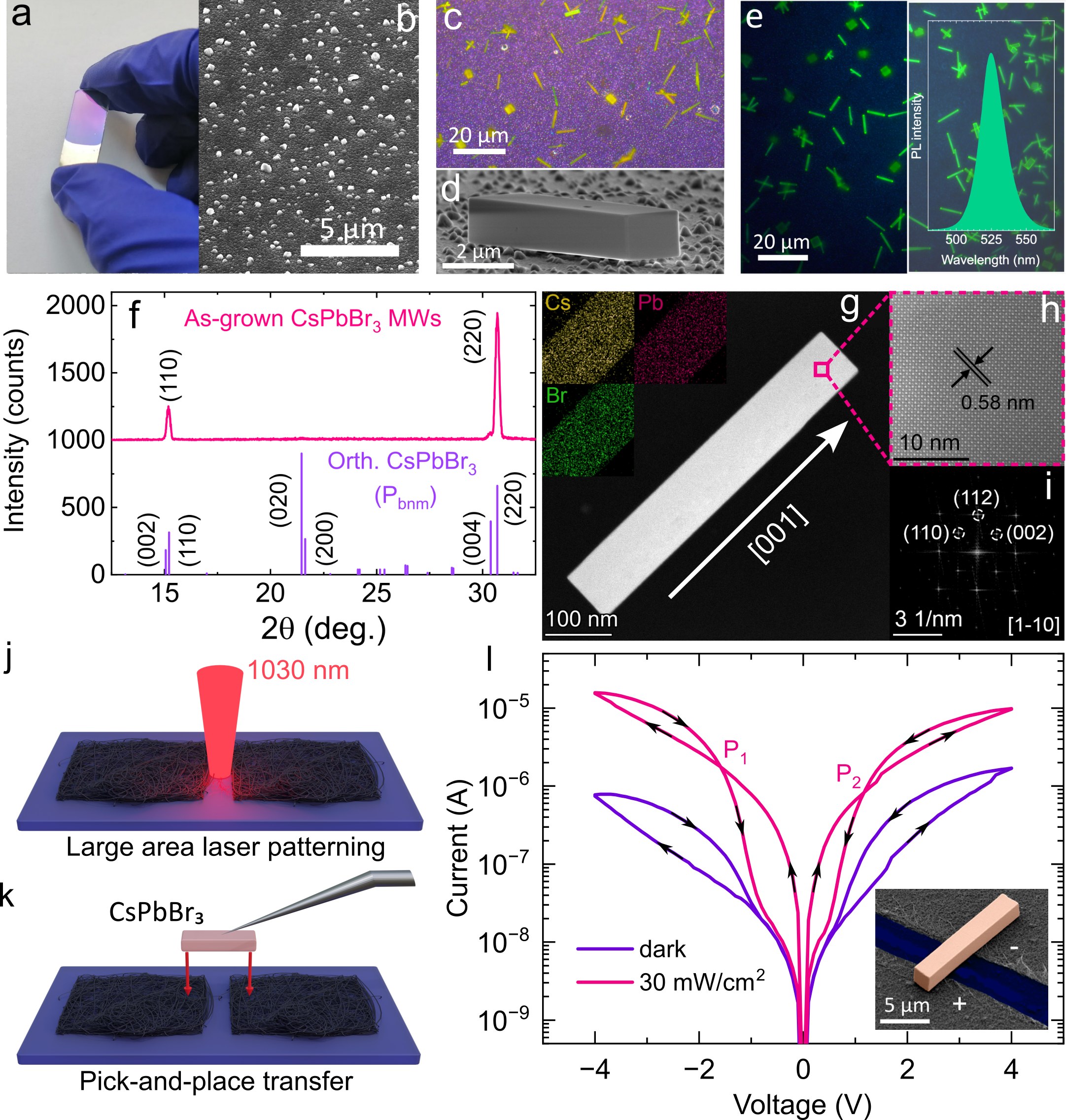}
	\caption{\textbf{Fabrication of perovskite microcrystals, characterization of their optical and structural properties, and manufacturing of a memristor device.} \textbf{(a)} Macroscopic image of the nanostructured Al$_2$O$_3$ substrate. \textbf{(b)} SEM-image of island-like morphology of the Al$_2$O$_3$ substrate. \textbf{(c)} Bright-field image of CsPbBr$_3$ microcrystals deposited on the substrate. \textbf{(d)} Tilted-angle SEM image of a single-crystal on top of Al$_2$O$_3$ islands. \textbf{(e)} Photoluminescent image and spectrum of the microcrystals. \textbf{(f)} XRD pattern for an ensemble of as-grown CsPbBr$_3$ microwire crystals and the reference pattern. \textbf{(g)} STEM image and EDX mapping of an isolated CsPbBr$_3$ nanowire crystal. \textbf{(h)} High-Resolution STEM image of a selected area \textcolor{black}{(indicated by a pink square on the (g)).} \textbf{(i)} FFT image obtained from (h). \textbf{(j)} Illustration of femtosecond laser cutting off SWCNT thin film into two symmetrical electrodes. \textbf{(k)} Dry transfer of a single crystal over SWCNT thin film electrodes. \textbf{(l)} Current-Voltage curve in the dark (purple curve) and under illumination (pink curve). $P_1$ and $P_2$ are points where the hysteresis type changes from the regular to the inverted one. Arrows indicate the voltage sweeping direction. The inset picture represents the SEM image of the as-fabricated device.}
	\label{fig:FIG.2}
\end{figure*}

\par The fabrication of the CsPbBr$_3$ single crystal memristor consists of the following stages. Firstly, we prepare nanostructured Al$_2$O$_3$ substrate with island-like morphology \cite{markina2023perovskite} (\textcolor{black}{Fig.\ref{fig:FIG.2} a,b; Fig.S1 a-e, SI}). Then we conduct a solution-processed synthesis similar to the previous work \cite{pushkarev2018few}. A small droplet (2 $\upmu$l) of metal halides (CsBr and PbBr$_2$) solution is dropped on the Al$_2$O$_3$ substrate and dried in the presence of \textcolor{black}{2-propanol-water azeotrope (H$_2$O$\cdot$IPA)} at 70 $^\circ$C for 15-20 min (for details, see the Methods Section). Thereafter, plenty of microcrystals are formed on the substrate (Fig.\ref{fig:FIG.2} c, Fig.S2 a-b, SI). \textcolor{black}{Scanning electron microscopy (SEM)} image shows that single crystals are deposited on top of islands and have regular rectangular shape (FIG.\ref{fig:FIG.2} d; Fig.S2 b, SI). The photoluminescence spectrum of the microcrystals contains a single peak centered at 524 nm (FIG.\ref{fig:FIG.2} e). X-ray diffraction (XRD) pattern collected from an ensemble of microwire crystals in $\theta-\theta$ geometry is represented by two peaks corresponding to crystallographic planes (110) and (220), that matches with the reference data for the orthorhombic phase (space group $P_{bnm}$)~\cite{rodova2003phase} and confirms their directional growth (FIG.\ref{fig:FIG.2} f). Scanning transmission electron microscopy (STEM) image of a nanowire crystal grown along [001] axis (FIG.\ref{fig:FIG.2} g) reveals its monocrystalline structure. Energy-dispersive X-ray spectroscopy (EDX) images demonstrate the uniform distribution of Cs, Pb, and Br in the nanowire (inserted image in FIG.\ref{fig:FIG.2} g). The distance between parallel planes is 0.58 nm which corresponds to crystallographic planes (002) of the orthorhombic phase of CsPbBr$_3$. Fast Fourier Transform (FFT) image obtained from Fig.\ref{fig:FIG.2} h contains sharp and bright points assigned to the planes (112), (110), and (002), indicating the periodic crystal lattice of CsPbBr$_3$ (FIG.\ref{fig:FIG.2} i). \textcolor{black}{To confirm the monocrystallinity of the perovskite microwire, selected area electron diffraction (SAED) was performed (see Fig. S3, SI). The SAED pattern in Figure S3b exhibits a series of sharp diffraction spots, including those corresponding to the (112), (110), and (002) planes.}

\par To fabricate the device electrodes, the single-walled carbon nanotube (SWCNT) thin film is dry-transferred \cite{nasibulin2011multifunctional} onto a flexible polyethylene naphthalate (PEN) substrate (\textcolor{black}{see Fig. S4, SI}). The SWCNT film is then patterned into two separate electrodes (Fig.\ref{fig:FIG.2} j)  by ablating a few-micrometer-wide path using a femtosecond laser \cite{marunchenko2022single} (for details, see the Methods Section; \textcolor{black}{Figs. S4, S5 and S6, SI}). To complete the device fabrication, the as-synthesized microwire crystal (Fig.\ref{fig:FIG.2} k) is dry-transferred onto the prepared SWCNT film electrodes (for details see Methods and \textcolor{black}{Figs. S7 and S8 in the SI}).

\par As a result, we obtain a lateral SWCNT-CsPbBr$_3$-SWCNT structure of \textcolor{black}{metal-semiconductor-metal (MSM)} type (Fig.\ref{fig:FIG.2} l; \textcolor{black}{Fig.S9, SI}), forming symmetrical Schottky contacts \cite{tao2019absolute,shiraishi2001work, marunchenko2023mixed}. The interelectrode distance of about 5 $\upmu$m (Fig.\ref{fig:FIG.2} l, inset image; \textcolor{black}{Figs. S6 and S9, SI}) is comparable with the charge carriers diffusion length \cite{stranks2013electron, dong2015electron, oksenberg2021deconvoluting} of the halide perovskite semiconductors which is desirable for efficient charge carrier extraction and subsequent improvement of the device performance. To validate the stability of the formed contact between SWCNT and CsPbBr$_3$, we applied optoelectrical stimuli to the device and measured the response for each of 5000 bending cycles. As a result, after 20000 bending cycles, the device response decreased to 85$\%$ of its initial value (Fig.S10, SI), thus manifesting the stable electrical contact of the as-formed structure. 

\par Current-voltage measurements on the device in the dark and under continuous wave (CW) light illumination of 30 mW/cm$^2$ at 532 nm wavelength are shown in Fig.\ref{fig:FIG.2} l (Figs. S11 and S12, SI; for a detailed description of the optoelectrical setup see Methods). The hysteresis $I-V$ curves in the dark have an inverted inductive-like response, where the current on the reverse scan is higher than that of the forward scan \cite{bisquert2021theory}. This behavior is characteristic of volatile memristors and is attributed to Schottky contacts modulated by mobile halide perovskite ions  \cite{marunchenko2023mixed, john2021halide, yen2021all, ravishankar2017surface}. 

\par Under light illumination the hysteresis becomes more complex (Fig.\ref{fig:FIG.2} l). At low voltages, it exhibits a capacitive-like response (where the current on the forward scan is higher than that of the reverse scan), while at higher voltages, it transitions to an inductive-like response. The transition between these two regimes is indicated on the plot as P$_1$ and P$_2$ points (around -1.6 V and 1.2 V, respectively). Such complex hysteresis transitions are commonly observed in halide perovskite optoelectronic devices with various structures \cite{munoz2022inductive, tress2015understanding, liu2019fundamental, berruet2022physical, rong2017tunable, gonzales2022transition, h2024accelerating}. Such hysteresis transition in Fig.\ref{fig:FIG.2} l  \cite{gonzales2022transition, berruet2022physical} is attributed to the coexistence of inductive and capacitive types of response in perovskite devices \cite{munoz2022inductive, bisquert2023transient}. It occures due to the combination of large surface capacitance of the perovskite, and the delay of recombination current, both due to ionic effects. Specifically, at low voltages, the polarization effect dominates, leading to capacitive behavior. However, at higher voltages, the electronic carriers cannot follow the ionic motion, giving rise to an inductive response. Therefore, our CsPbBr$_3$ MSM device contains at least two competing processes that can be qualitatively described as capacitive and inductive circuit branches (see SI Fig.S13). 

\textcolor{black}{Building on this observation, we further  investigated how the voltage scan rate affects the hysteresis transition points. Notably, increasing the scan rate shifts the hysteresis transition points to higher (absolute) voltages. In Figure S14, we show that for a similar sample, increasing the scan rate from 0.5 V/s to 10 V/s shifts the hysteresis transition voltage from 2 V to over 3 V. Beyond 60 V/s, the hysteresis undergoes a transformation: the transition from capacitive-like to inductive-like behavior disappears, and the device exhibits a purely capacitive-like response. At even higher scan rates, the hysteresis gradually vanishes and eventually becomes purely inductive-like. These results show that the conductive and inductive behavior of the device can be easily controlled by adjusting the voltage scan rate, allowing for tunable hysteresis responses.}

\begin{figure*}[t]
	\centering
	\includegraphics[scale = 0.8]{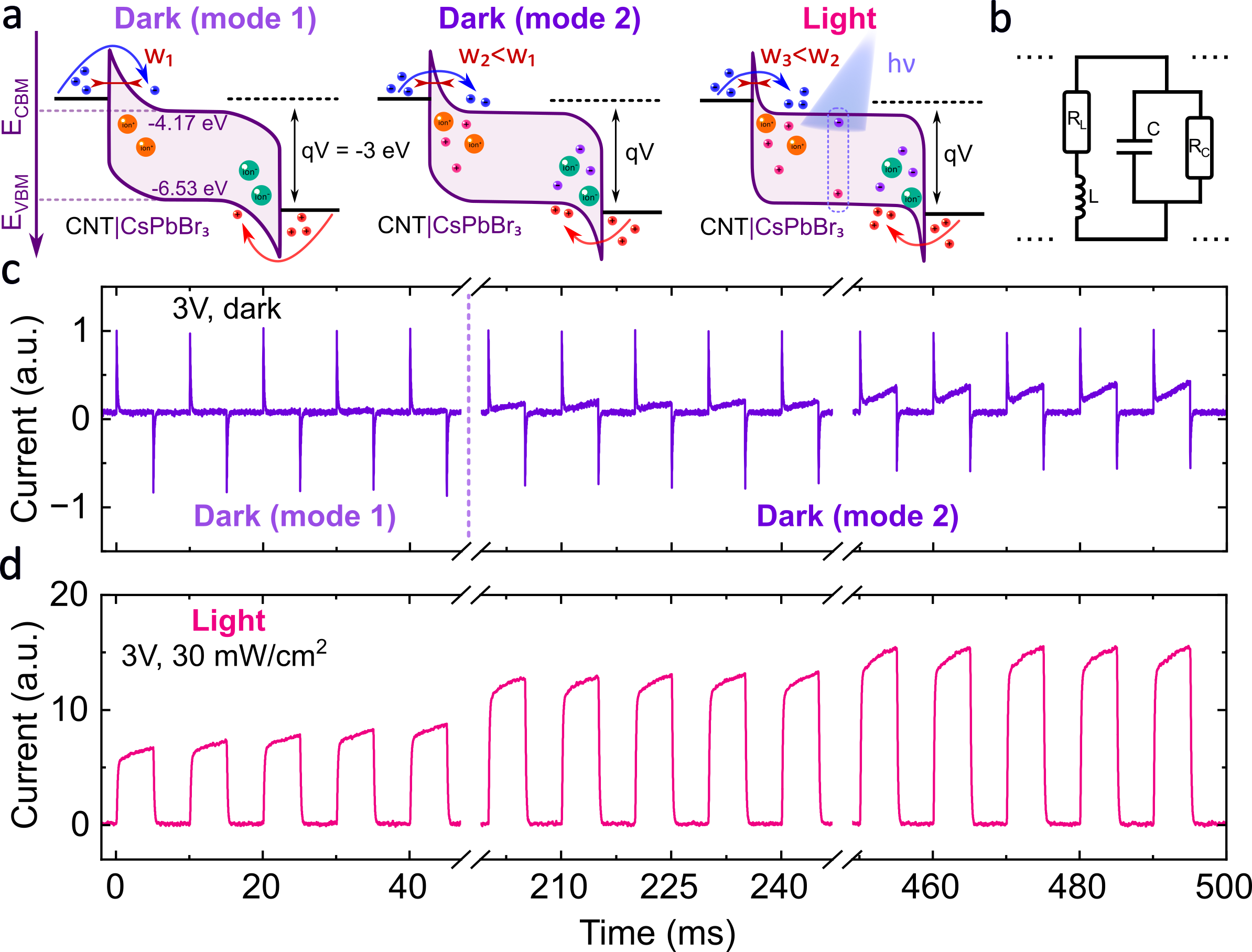}
	\caption{\textbf{Transient dynamics and equivalent scheme approach.} \textbf{(a)} The schematic electronic band diagrams are the corresponding parts of the electrical circuit equivalent scheme \textbf{(b)} and type of the transient dynamics. \textcolor{black}{Mode 1 in the dark case corresponds to initial pure capacitive regime which transforms to mode 2, when ionic component starts growing. Under hybrid optoelectrical stimuli for 3V pure inductive regime is implemented.} On the diagrams big orange and green points represent intrinsic perovskite mobile ions, coexisting together with injected electrons (blue) and holes (red), and with photogenerated electrons and holes in case of light illumination. \textcolor{black}{W$_i$ denotes the width of Schottky barrier}. The transient dynamics (3 V, 5 ms pulses) of perovskite memristor in the dark case \textbf{(c)} and under hybrid optoelectrical stimuli \textbf{(d)}, respectively. The current is normalized to the maximum value of dark current.}
	\label{fig:FIG.3}
\end{figure*}

\par To explain the transient dynamics of the competing dynamic processes we provide the schematic electronic band diagrams of the device and the equivalent electrical circuit. The memristive behavior of our device is attributed to the modulated Schottky contacts by the intrinsic perovskite ions, such as bromide ions and their vacancies \cite{marunchenko2023mixed}. The accumulation of the mobile ions at the interfaces leads to electrostatic lowering and shrinking of the energy barrier at the carbon nanotube-perovskite interface (Fig.3 a). 
In the equivalent circuit this can be represented by the inductive branch (resistor and inductor in series). Because the inductive component is attributed to the slowly moving ions, their appearance in the transient dynamics can be much slower, compared to the capacitive response of the electronic nature. Following the application of the 3V bias one can see the slow change from the capacitive to inductive dynamics, manifesting the change of the Schottky barrier (Fig.3 c) \cite{gonzales2024capacitive}. The parallel connection of the inductive and capacitive circuit branches highlights the competition between two contributions to the transient current response (Fig.3 b) (see also Fig.S13, SI). A similar scheme has been successfully validated in the literature describing halide perovskite optoelectronic devices \cite{rong2017tunable, munoz2022inductive, filipoiu2022capacitive, gonzales2022transition, berruet2022physical, bisquert2023resistance, bisquert2021theory, bisquert2022chemical}. 

\par Under light illumination many additional processes may be activated in halide perovskites. It includes the generation of photocarriers, carrier trapping, heating, chemical reactions etc. \cite{marunchenko2024hidden, marunchenko2024charge, motti2019defect}. The huge modulation of the current across the device under light illumination (Fig.2 l), means that the photoconductive effect dominates the transient response of the device under hybrid optoelectrical stimuli (Fig.3 a) \cite{marunchenko2022single}. Here, the nature of the Schottky interface allows the photogenerated carriers to additionally contribute to the modulation of the Schottky barrier \cite{marunchenko2023mixed} thus increasing the overall current and resulting in the inductive-like behavior of the device (Fig.3 d). 

\subsection{Short-term BCM response}

\par
\begin{figure*}[t]
	\centering
	\includegraphics[scale = 0.8]{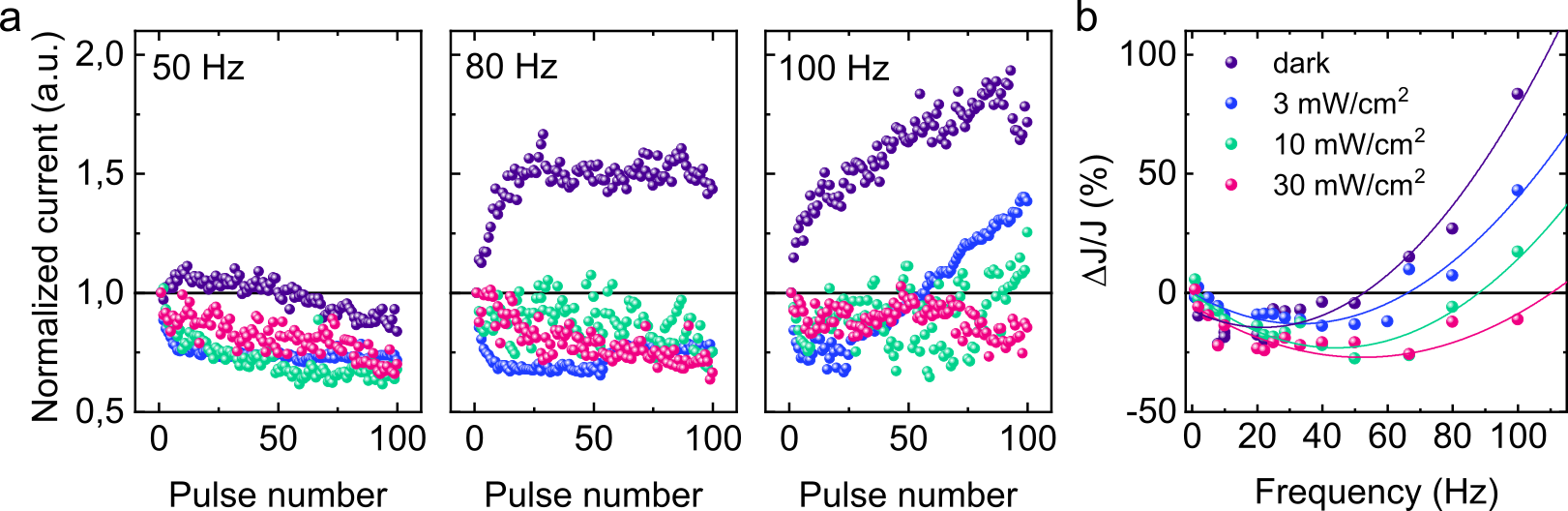}
	\caption{\textbf{Experience-dependent synaptic plasticity featuring short-term Bienenstock-Cooper-Munro learning under hybrid optoelectrical stimuli.} \textbf{(a)} Current response (normalized to the initial current) to 100 consecutive pulses (2.5 V, 5 ms) with different frequencies under electrical and optoelectrical stimuli. \textbf{(b)} The synaptic weight change inside a sequence of 100 pulses versus their frequency shows the "tick-like" shape similar to the BCM theory. The plot is separated into two parts for positive and negative relative change in synaptic weight. The light intensity shifts the zero point of the curve, showing the sliding threshold property. The solid lines indicate parabolic fit.}
	\label{fig:FIG.4}
\end{figure*}

 For a deeper investigation of the dynamical synaptic behavior of our memristors, we applied trains of 100 voltage pulses with an amplitude of 2.5 V and a pulse width of 5 ms. We monitored the current response as a function of pulse train frequency and light illumination intensity (Fig.\ref{fig:FIG.4} a). We observed that the current evolves non-monotonically over the course of 100 voltage pulses. 

\par In the dark, at high frequencies (100 Hz), the current initially rises rapidly, following an inductive-like transient response, before transitioning into a slower increase. At intermediate frequencies (80 Hz), the current stabilizes, whereas at low frequencies (50 Hz), it may even decrease. Under varying light intensities (3 mW/cm$^2$ - 30 mW/cm$^2$), the current exhibits different behaviors: at low frequencies, it decreases over the pulse train, indicating capacitive-like behavior, whereas at higher frequencies, it stabilizes or undergoes a steep increase. This inductive-like behavior can be attributed to the activation of ionic transport, which occurs alongside the relatively faster photogenerated electronic transport, also known as photoinduced capacitance. 

\textcolor{black} {It is important to note that the observed behavior, particularly following the +2.5 V pulse train at 100 Hz under illumination with the light power density of 30 mW/cm$^2$, which exhibits a capacitive-like transient response, is qualitatively different from that demonstrated in Figure 3d, where a similar +3 V pulse train results in inductive-like behavior. Despite this difference, in the I–V curve (Fig. \ref{fig:FIG.2}l), both +2.5 V and +3 V fall within the same inductive-like branch. 
This counterintuitive discrepancy can be attributed to the significantly higher voltage scan rates in pulsed mode compared to quasi-DC I–V measurements in Figure 2l. As mentioned previously, higher voltage scan rates shift the hysteresis transition point toward higher voltages (Fig. S14, SI), which we believe accounts for this effect. Such a sensitive transition between capacitive and inductive behavior can be leveraged for neuromorphic functional systems, as will be demonstrated below.} 

\par For a clearer comparison between different light intensities for each pulse train, we track the relative current change over the pulse train, calculated as:

\begin{equation*}
    \Delta J/J = \left(\langle J_{\text{last}}\rangle-\langle J_{\text{first}}\rangle \right)/\langle J_{\text{first}}\rangle,
\end{equation*}

where $\langle J_\text{first(last)}\rangle$ represents the current averaged over the first (last) five pulses, plotted as a function of train frequency for various light intensities (Fig.\ref{fig:FIG.4} b). In our case $\Delta J/J$ serves as the measure of short-term synaptic plasticity of the optoelectronic memristor, governing the assignment of potentiation and depression regions to the areas above and below $\Delta J/J$=0, respectively. 

\par For our single crystal memristors we observe that for each light intensity the $\Delta J/J(f)$ dependence follows the same trend: starting from 0 (which corresponds to the idle situation, obviously, not leading to any synaptic modification), the relative current change is negative and further decreases with frequency increase, reaching some minimum value and turning to the monotonic growth. Importantly, the value of the so-called threshold frequency (at which $\Delta J/J$ is equal or interpolated to 0) increases monotonically with the light intensity (Fig.S15). These findings closely align with the BCM theory of synaptic plasticity in the visual cortex \cite{bienenstock1982theory, wang2020toward, huang2021memristor}. Since BCM plasticity plays a crucial role in visual pattern recognition, its direct integration into optoelectronic neuromorphic devices is highly valuable. \textcolor{black}{Importantly, the reported device dynamics falls within a timescale range from milliseconds to seconds (Fig. 3c,d; Fig. 4a,b), thus aligning with biological timescales. Moreover, the studied light intensities correspond to natural visible light conditions, spanning from overcast or shaded environments to bright, direct sunlight.}

\par The effects of light illumination on the $\Delta J/J(f)$ dependence – namely, the increase of threshold frequency (Fig.\ref{fig:FIG.4} b) and the decrease in $\Delta J/J$ value at high frequencies – are analogous to the experience-dependent plasticity described by the BCM theory. In neurobiological systems, sensory deprivation during an organism’s lifetime results in a sliding threshold frequency and increased electrical sensitivity in brain regions responsible for visual information processing. 
\par Notably, in our memristors, short-term BCM behavior is implemented directly, replicating classical neurobiological experiments without requiring complex preconditioning techniques such as preliminary electrical excitation at different frequencies \cite{du2015biorealistic, khanas2022second}, \textcolor{black}{\cite{zhu2023implementation}, \cite{ke2022bcm}, \cite{chen2024proton}, \cite{guo2020bienenstock}} or triplet STDP \cite{wang2020toward, john2022ionic}. This simplicity is advantageous for technological applications. \textcolor{black}{The lack of long-term memory in our memristor does not reduce its potential for neuromorphic computing. In fact, its simple rate-based coding makes it well-suited for the use in a dedicated neuromorphic core that handles complex plasticity rules. This allows for faster synaptic weight updates, which can be stored in external non-volatile memristor cells, following the SpiNNaker architecture \cite{galluppi2015framework} and IBM’s PCM-based crossbar arrays \cite{ambrogio2018equivalent}.}

\section{Conclusion}
\par \textcolor{black}{In summary, we have investigated single-crystal CsPbBr$_3$ memristors assembled in a lateral geometry on a flexible substrate. These devices exhibit highly responsive photodetection (800 A/W at 2.5 V of applied bias and under 30 mW/cm$^2$ illumination) while maintaining the performance under the mechanical deformation of 20000 cycles. Our devices exhibit biorealistic synaptic behavior, where light intensity dynamically adjusts the sliding frequency threshold in line with Bienenstock-Cooper-Munro theory. Importantly, the BCM functionality is achieved within a biologically relevant light intensity range of 1 mW/cm$^2$ - 30 mW/cm$^2$, aligning well with natural visual processing conditions. Furthermore, the modulation frequency of the BCM behavior spans a range of 0 Hz to 100 Hz at 2.5 V of applied bias, mirroring the timescales of the biological systems. These quantitative benchmarks establish our memristors as a promising platform for neuromorphic computing applications.}
\par By integrating retinomorphic (photodetecting) properties with visual cortex-inspired BCM synaptic behavior within a single halide perovskite microcrystal memristor, we introduce a material platform that could streamline visual information processing. Unlike both biological and conventional electronic systems, where sensing (photoreceptors/cameras), information transfer (neurons/interfaces), and processing (visual cortex/software) are distinctly separate, our approach suggests a more compact and efficient alternative. The in-sensor computing capabilities demonstrated in this work highlight the potential for seamless integration into neuromorphic chips for real-time visual data processing, paving the way for energy-efficient, hardware-based vision systems with reduced computational complexity.

\section{Methods}

\subsection{Fabrication of substrates with island-like morphology}

\textcolor{black}{To fabricate Al$_2$O$_3$ substrates with island-like morphology we utilized a protocol similar to previously reported one~\cite{markina2023perovskite}. First, a thin aluminum film was thermally deposited under vacuum on glass substrates. Then, substrates were immersed in a 0.3 M aqueous solution of oxalic acid and anodized at 7 V for 15 minutes. After anodizing, metallic film transformed into a non-conductive Al$_2$O$_3$ layer (Fig. \ref{fig:FIG.2} a,b; Fig. S1, SI). Finally, substrates were rinsed with deionized water to wash away acid moieties.}

\subsection{Synthesis of CsPbBr$_3$ microcrystals}

\textcolor{black}{To obtain CsPbBr$_3$ microcrystals we employed a wet chemical synthesis using a protocol similar to one reported by Pushkarev et al. \cite{pushkarev2018few}. Firstly, we prepared 0.1 M solution of stoichiometrically mixed metal halides (PbBr$_2$, 99.999$\%$, Alfa Aesar, and CsBr, 99.99$\%$, Sigma-Aldrich) in anhydrous dimethyl sulfoxide (DMSO, 99.8$\%$, Alfa Aesar) in a N$_2$-filled glovebox. 2 $\upmu$L of the solution was drop-casted onto the Al$_2$O$_3$ substrate at ambient conditions and then the substrate was sealed in a hot Petri dish containing 250 $\upmu$L of H$_2$O$\cdot$IPA azeotrope (95$\%$) and dried on a hotplate at 70~$^\circ$C for 15 min. As a result, numerous microcrystals distributed over the substrate with a mean density of 9.6$\times$10$^5$~cm$^{-2}$ were grown and further utilized for a dry transfer procedure (Fig. S2, SI).} 

\subsection{Synthesis of SWCNT}

Thin films of single-walled carbon nanotubes (SWCNTs) were produced using an aerosol (floating catalyst) chemical vapor deposition (CVD) method \cite{khabushev2019machine} followed by a simple filtration via a nitrocellulose membrane (HAWP, Merck Millipore). For the aerosol CVD, ferrocene (98\%, Sigma Aldrich) was employed as a catalyst precursor while CO (99.99\% Linde gas) acted as a carbon source. Ferrocene was transferred with a CO stream as a vapor to decompose upon reaching the hot zone (880 $^\circ$C) of a quartz tube. The formation of Fe nanoparticles from ferrocene allows the catalytic Boudouard reaction (2CO = C + CO$_2$) to nucleate and produce nanotubes. Carbon dioxide (99.995\% MGPZ) introduction tunes the nanotube properties (e.g., diameter distribution) and governs the process reactor productivity \cite{khabushev2020structure}. The collected films consisted of individual SWCNTs exhibiting a mean diameter of ca. 1.9 nm, high quality, and random spatial orientation. After the deposition, SWCNTs form a randomly oriented uniform film suitable for further dry-transfer \cite{kaskela2010aerosol} onto different types of a substrate \cite{gilshteyn2019mechanically}.

\subsection{Device fabrication}
\textit{SWCNT electrode preparation.}~\textcolor{black}{The process of making electrodes from a SWCNT film consists of two stages. At the first stage, the film of 20 nm thickness on a cellulose filter paper was dry transferred to a polyethylene naphthalate (PEN) substrate cleaned with H$_2$O$\cdot$IPA. After the removal of the filter paper, 20 $\upmu$l of H$_2$O$\cdot$IPA was dripped to the transferred film to improve its adhesion to the substrate. Then, two droplets of silver paste (RS 186-3600) were deposited on the film and dried at the temperature $T$=70 $^\circ$C for 20 min. The droplets act as contact pads for electrical connection (Figs. S4 and S16, SI). Then, the sample was stuck to double side tape on a glass substrate for further processing (Fig. S4, SI).} 

\textcolor{black}{At the second stage, laser ablation of the film to form interelectrode space was carried out using a fs laser (AVESTA ANTAUS-10W-4u/2.5M) with a pulse length of 270 fs, wavelength of 1030 nm, and pulse repetition frequency 4 kHz. Laser radiation with fluence of $\sim$ 2.3$\pm$0.5 J/cm$^2$ was focused through a 50$\times$ objective (NIR Mitutoyo, NA = 0.42) on the sample surface at normal incidence (see Figs S5, and S6, SI).}

\textit{Dry transfer of microwires.}~\textcolor{black}{Perovskite microcrystals on the Al$_2$O$_3$ substrate can be dry transferred to the SWCNT electrodes by using an adhesive polydimethylsiloxane (PDMS) lens. The lens was prepared according to the following protocol. A silicon elastomer base (SYLGARD 184) and a curing agent (SYLGARD 184) were mixed in a 2:1 ratio by mass and stirred vigorously (see Fig. S7 (a), SI). The resultant mixture was further degassed (see Fig. S7(b), SI), and a droplet of the mixture was deposited onto a glass substrate and kept at T=90 °C in an oven for 1 h. Note the glass substrate with the droplet must be twisted upside down during the curing process to give a semi-convex shape of the lens that makes it possible to detach a single perovskite microcrystal from the Al$_2$O$_3$ substrate. The droplet was then completely solidified at room temperature for 12 h.}

\textcolor{black}{Thereafter, the prepared substrate with the PDMS lens was mounted on a three-axis positioner on an optical microscope (Zeiss Axio imager 2). The dry transfer procedure illustrated in Figure S8 (see
SI) was conducted under 20$\times$ objective (EC Epiplan-Neofluar 20x/0.50).}


\subsection{Optoelectrical measurements}

To perform combined optoelectrical measurements the sample was mounted on Cascade Microtech Summit 11000M probe station. The probes were landed on Pt layers deposited onto two separated Si substrates, which in turn were connected with the SWCNT thin film electrodes via 50 $\upmu$m gold wires and silver paste. Keysight B1500A semiconductor device analyzer was used for the electrical measurements. The optical stimuli were applied using CW 532 nm laser with the power density of 30 mW/cm${^2}$. I-V curves were measured using the source-measure unit (SMU) of the Keysight B1500A semiconductor device analyzer in the quasi-DC mode with the staircase sweeps with minimum integration time per voltage point of 3 ms. Pulse measurements for BCM behavior demonstration were carried out using Keysight B1530A waveform generator/fast measurement unit (WGFMU) \textcolor{black}{(see Fig. S16, SI).} \textcolor{black}{All measurements were performed at room temperature.}

\subsection{\textcolor{black}{Contact stability measurements}}

\textcolor{black}{To perform a contact stability test a bending machine similar to one described in the work \cite{kasewieter2013failure} was used. The substrate with a perovskite memristor was located on the lower cylinder, while the two upper ones bent and unbent it in the direction shown by the arrows in Figure S10 (see SI). Curvature radius of the bent sample is $R = $ 2 mm.} 

\textcolor{black}{After every 5000 bending cycles, the sample was removed from the setup and examined on optoelectrical circuit consisting of wave generator (Keysight 33600A), oscilloscope (Keysight DSOX6004A), and CW laser (Fianium Supercontinuum). The sample was then electrically excited by 100 voltage pulses with the width of 5 $\upmu$s and the amplitude of 3 V upon illumination by the CW laser with wavelength of 532 nm and intensity of 30 mW/cm$^2$. The response (photocurrent) of the sample to this excitation measured through the shunt resistance $R = 1$ MOhm is presented in Figure S10a (see SI). Subsequently, the dependence of the maximum current value on the number of bending cycles was recorded and depicted in Figure S10b.}

\medskip
\section*{Supporting Information} \par
Supporting Information is available from the Wiley Online Library or from the author.

\medskip
\section*{Acknowledgements} \par 
\textcolor{black}{This work was supported by the Russian Science Foundation (project No. 23-72-00031).} The authors gratefully acknowledge the Shared Research Facilities Center of Moscow Institute of Physics and Technology (SRF MIPT) for the access to the probe station and SEM. A.G.N. acknowledges the Russian Science Foundation (project No 22-13-00436) for the synthesis of carbon nanotubes. The authors thankfully acknowledge the ITMO-MIPT-Skoltech Clover Program.

\section*{Author contribution}

A.M., A.K. and I.M. originated the idea. A.G.N. and D.V.K. synthesized single-walled carbon nanotube thin films. R.P., I.M., S.A. and A.E. contributed to the fabrication of the memristors. I.M., A.K., A.M., N.S. performed optoelectrical measurements of memristors. A.M, I.M., A.Z., A.K., D.S., A.Y. analyzed the data. A.P., A.Z., J.B., and A.M. supervised the project. A.M., I.M., A.K and A.P. wrote the original draft. A.M. managed the whole project. All authors reviewed and edited the manuscript. All authors contributed to the discussions and commented on the paper.

\medskip
\section*{Conflict of Interest}

The authors declare no conflict of interest.

\medskip

\bibliography{main.bib}
\bibliographystyle{MSP}

\clearpage

\begin{figure}[t]
\centering
\textbf{Table of Contents}\\
\medskip
  \includegraphics[scale = 1]{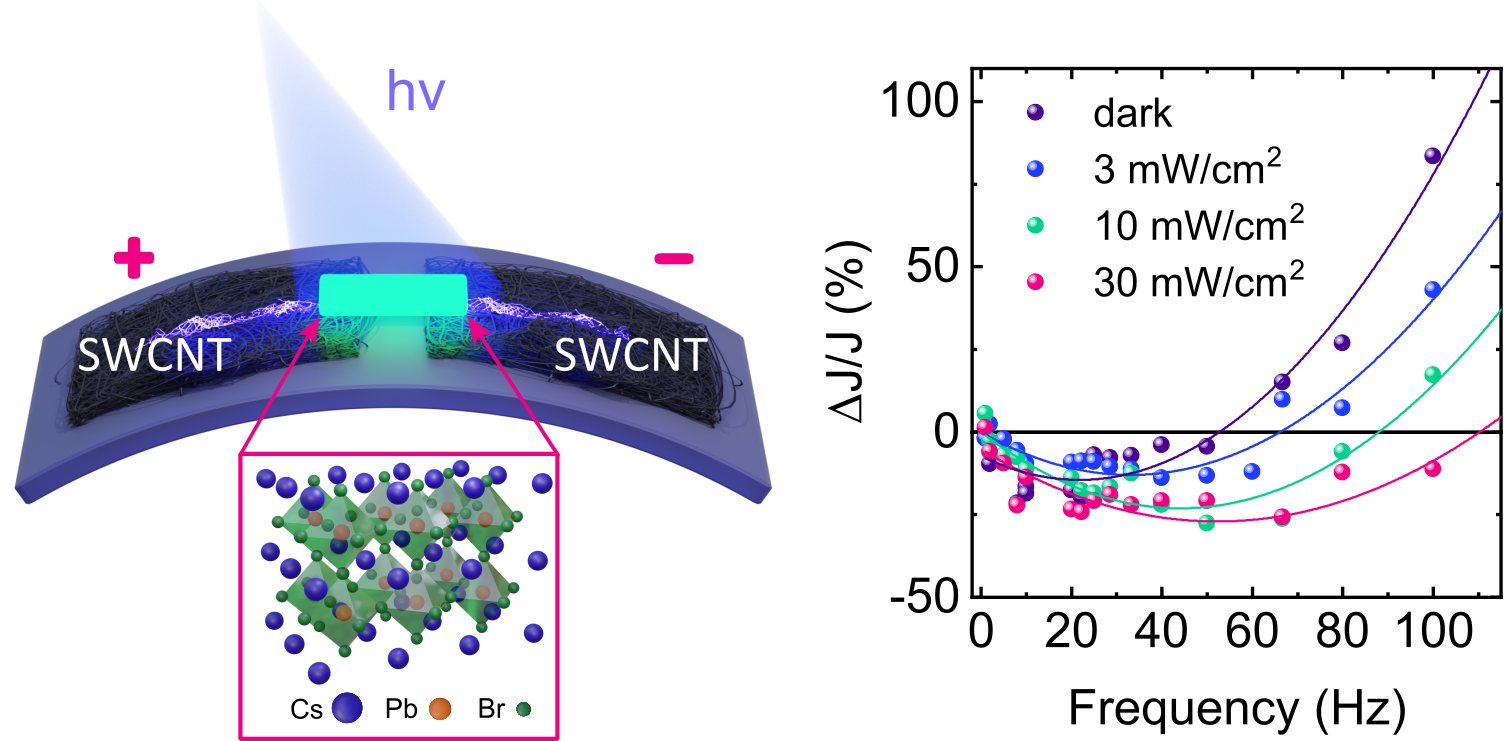}
  \medskip
    \caption*{We present flexible CsPbBr$_3$ perovskite memristors with retinomorphic photodetection and short-term Bienenstock-Cooper-Munro synaptic behavior. These devices hold promise as a compact, energy-efficient alternative for neuromorphic vision systems.}
\end{figure}

\end{document}


\pagestyle{fancy}
\rhead{\includegraphics[width=2.5cm]{vch-logo.png}}
\newcommand{\red}{\textcolor{red}}
\newcommand{\blue}{\textcolor{black}}

\renewcommand{\thefigure}{S\arabic{figure}}	
	\renewcommand{\theequation}{S\arabic{equation}}	
	\renewcommand{\thesection}{S\arabic{section}.}

\title{Supplementary for: Short-Term Bienenstock-Cooper-Munro Learning in Optoelectrically-Driven Flexible Halide Perovskite Single Crystal Memristors} 

\maketitle


\author{Ivan Matchenya, $^{1,2}$}
\author{Anton Khanas, $^{3}$}
\author{Roman Podgornyi, $^{1}$}
\author{Daniil Shirkin, $^{1}$}
\author{Alexei Ekgardt, $^{1}$}
\author{Nikita Sizykh, $^{3}$}
\author{Sergey Anoshkin, $^{1}$}
\author{Dmitry V. Krasnikov, $^{2}$}
\author{Alexei Yulin, $^{1}$}
\author{Alexey Zhukov, $^{4}$}
\author{Albert G. Nasibulin, $^{2}$}
\author{Ivan Scheblykin, $^{5}$}
\author{Anatoly Pushkarev, $^{2,*}$}
\author{Andrei Zenkevich, $^{3,*}$}
\author{Juan Bisquert, $^{6,*}$}
\author{Alexandr Marunchenko, $^{1,5,*}$}


\begin{affiliations}
$^1$ ITMO University, School of Physics and Engineering, St. Petersburg, 197101, Russian Federation

$^2$ Skolkovo Institute of Science and Technology, 30/1 Bolshoy Boulevard, 121205 Moscow, Russian Federation

$^3$ Moscow Institute of Physics and Technology (National research university), Institutskiy per.
9, Dolgoprudny, Moscow Region, 141701, Russia

$^4$ International Laboratory of Quantum Optoelectronics, HSE University, 190008, St. Petersburg, Russian Federation

$^5$ Chemical Physics and NanoLund, Lund University, P.O. Box 124, 22100 Lund, Sweden

$^6$ Instituto deTecnologia Quimica (Universitat Politei cnica de Valencia-Agencia Estatal Consejo Superior de Investigaciones Cientificas), 46022 Valencia, Spain

Andrey Zenkevich\\
Email Address:
zenkevich.av@mipt.ru

A.P. Pushkarev\\
Email Address:
anatoly.pushkarev@metalab.ifmo.ru

Juan Bisquert\\
Email Address:
bisquert@uji.es\\

A.A. Marunchenko\\
Email Address:
a.marunchenko@metalab.ifmo.ru\\

\end{affiliations}

\makeatletter
\renewcommand{\@maketitle}{%
{%
\thispagestyle{empty}%
\vskip-36pt%
{\raggedright\sffamily\bfseries\fontsize{20}{25}\selectfont \@title\par}%
\vskip10pt
{\raggedright\sffamily\fontsize{12}{16}\selectfont  \@author\par}
\vskip25pt%
}%
}%
\makeatother

\newpage

\begin{figure}
    \centering
    \includegraphics[width=0.75\linewidth]{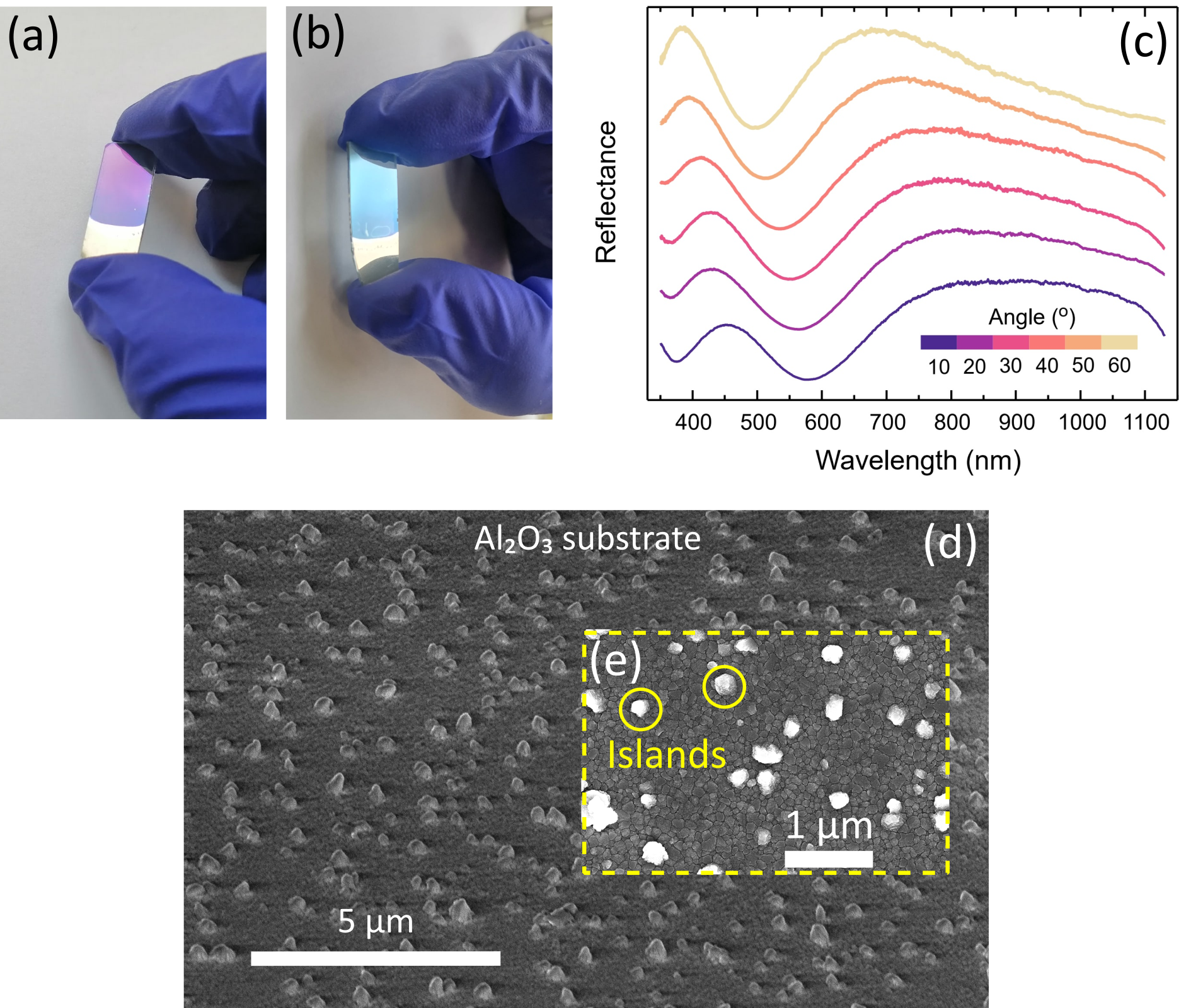}
    \caption{\textcolor{black}{\textbf{Characterization of nanostructured Al$_2$O$_3$ substrates.} \textbf{(a-b)} Images of the nanostructured Al$_2$O$_3$ substrate with island-like morphology. \textbf{(c)} Reflectance spectra of the Al$_2$O$_3$ substrate at different angles. SEM image \textbf{d} and HRSEM image \textbf{(e)} of the Al$_2$O$_3$ substrate higlighting formation of islands.}}
    \label{fig:S1}
\end{figure}

\begin{figure*}
	\centering
	\includegraphics[scale = 0.9]{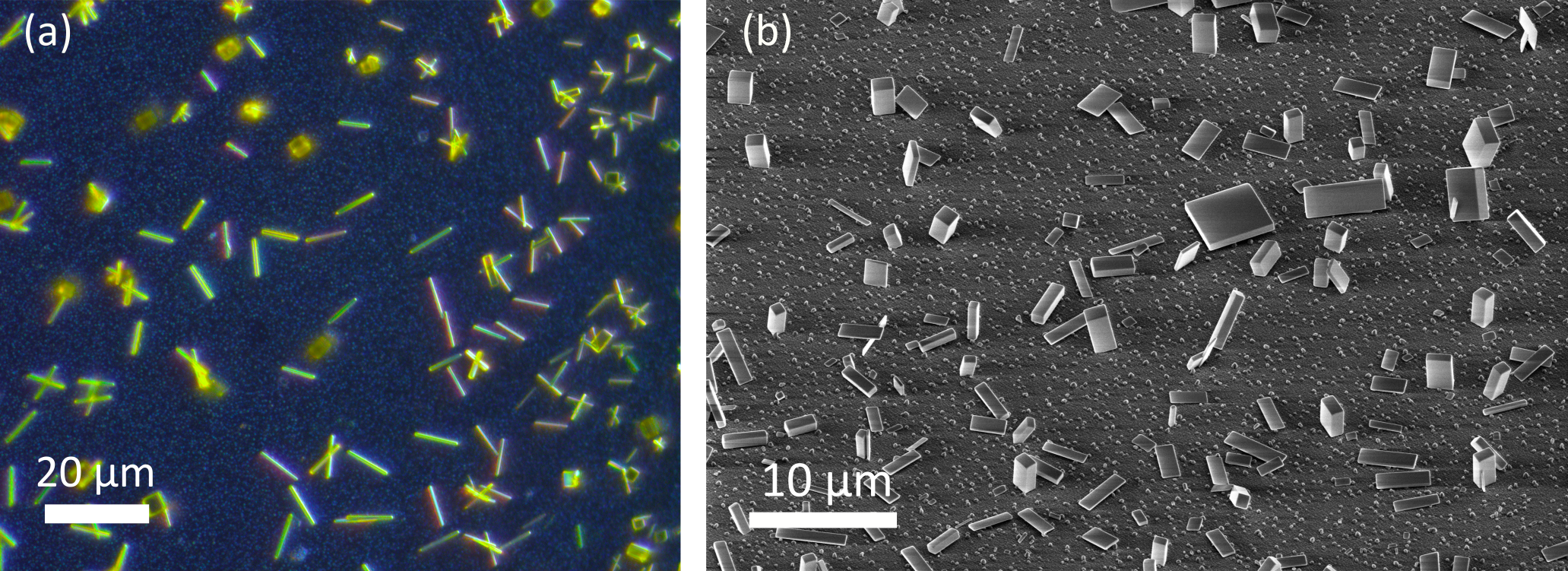}
	\caption{\textbf{Synthesis of the CsPbBr$_3$ microcrystals.} \textbf{(a-b)} Dark-field and SEM image of microcrystals grown on the Al$_2$O$_3$ substrate}
	\label{fig:S2}
\end{figure*}

\newpage

\begin{figure*}
	\centering
	\includegraphics[scale = 0.5]{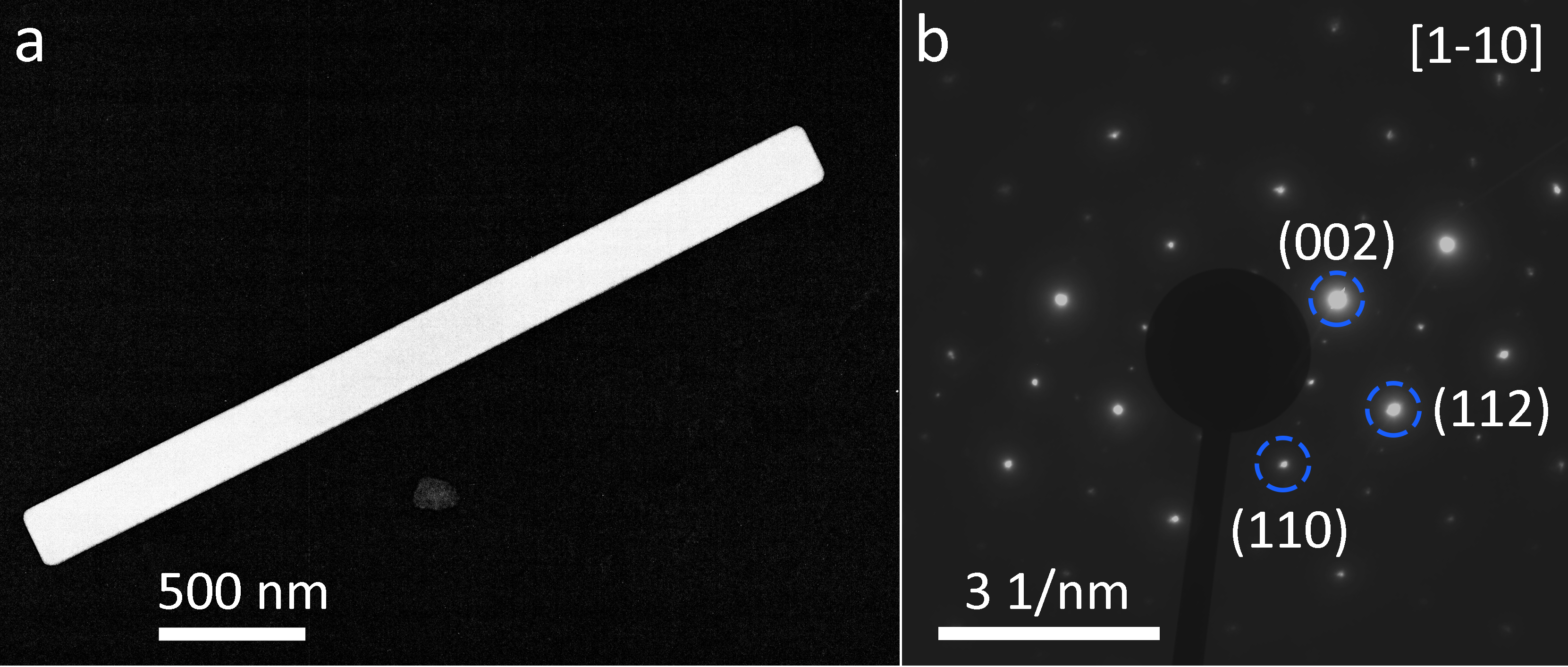}
	\caption{\textcolor{black}{\textbf{Establishing the monocrystallinity of a single perovskite microcrystal.}  \textbf{(a)} Low resolution HAADF-STEM image of a single microwire. \textbf{(b)} SAED image of the entire microwire revealing sharp diffraction spots assigned to planes of perovskite lattice and confirming the microwire monocrystallinity}}
	\label{fig:S3}
\end{figure*}

\begin{figure*}
	\centering
	\includegraphics[scale = 0.85]{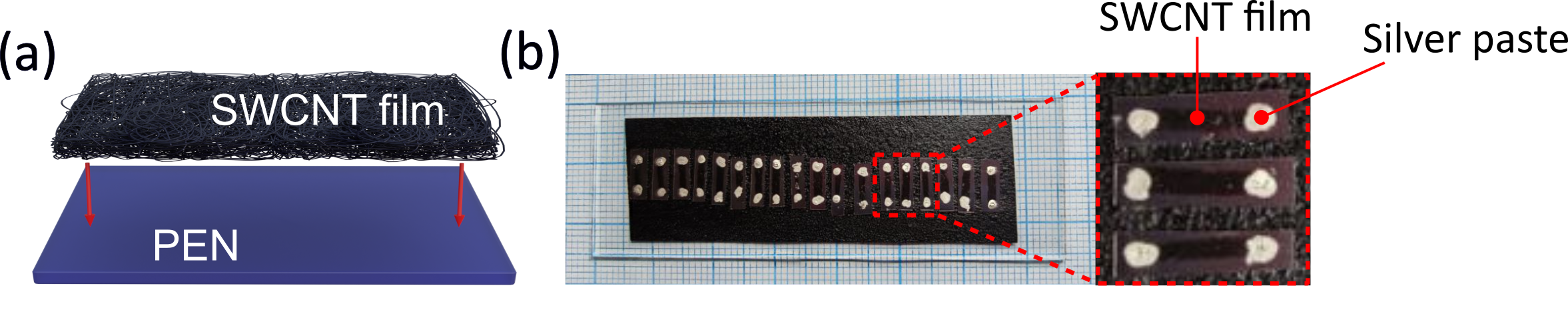}
	\caption{\textcolor{black}{\textbf{} (a) Transfer of SWCNT film to the PEN substrate, (b) prepared PEN substrate with transferred SWCNT film and droplets of silver paste for laser ablation}}
	\label{fig:S3}
\end{figure*}

\begin{figure*}
	\centering
	\includegraphics[scale = 0.85]{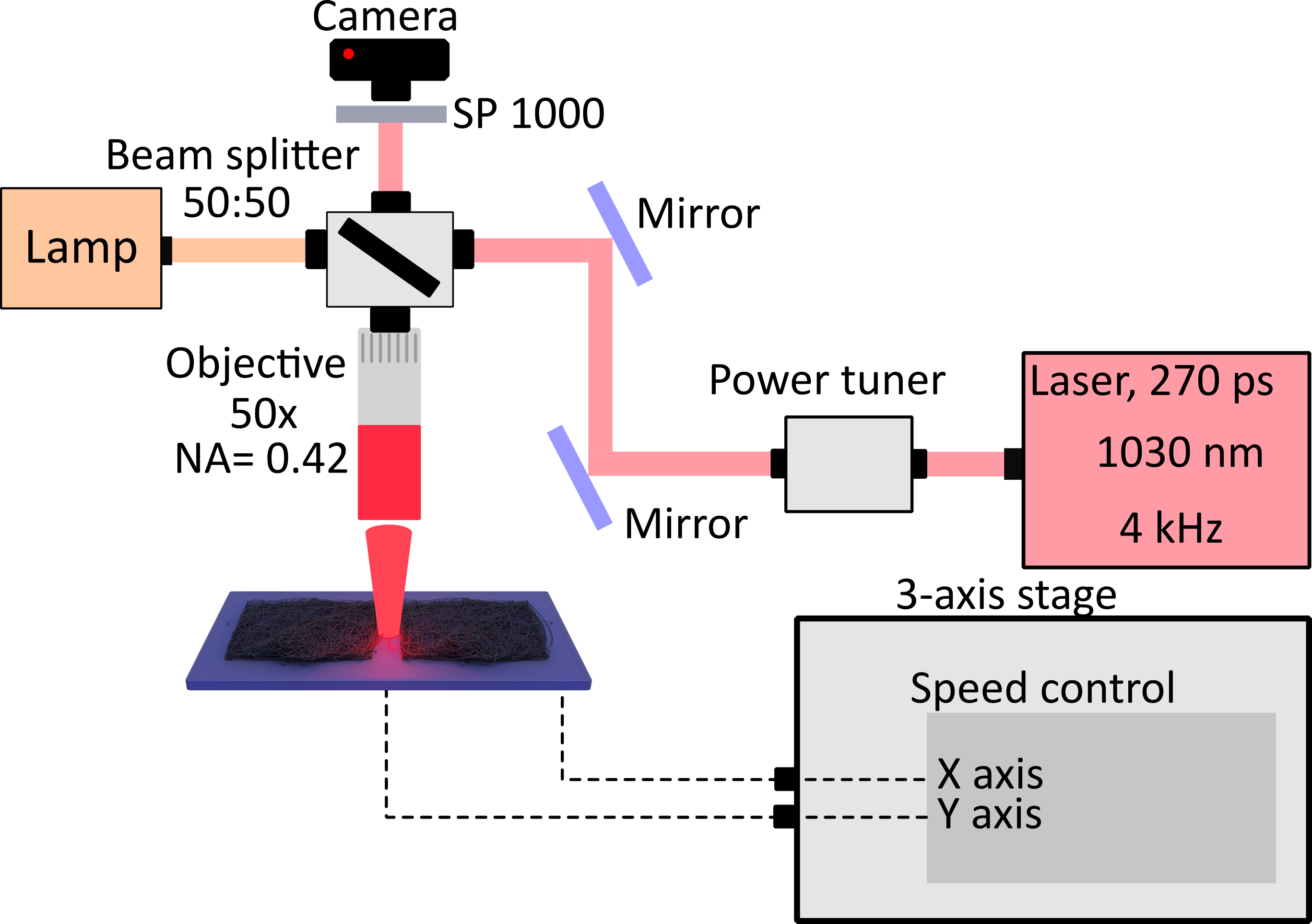}
	\caption{\textcolor{black}{\textbf{Optical scheme for laser ablation.} The velocity and acceleration of the Standa 3-axis stage were chosen in such a way that there were no short circuits of SWCNT.}}
	\label{fig:S4}
\end{figure*}

\begin{figure*}
	\centering
	\includegraphics[scale = 0.85]{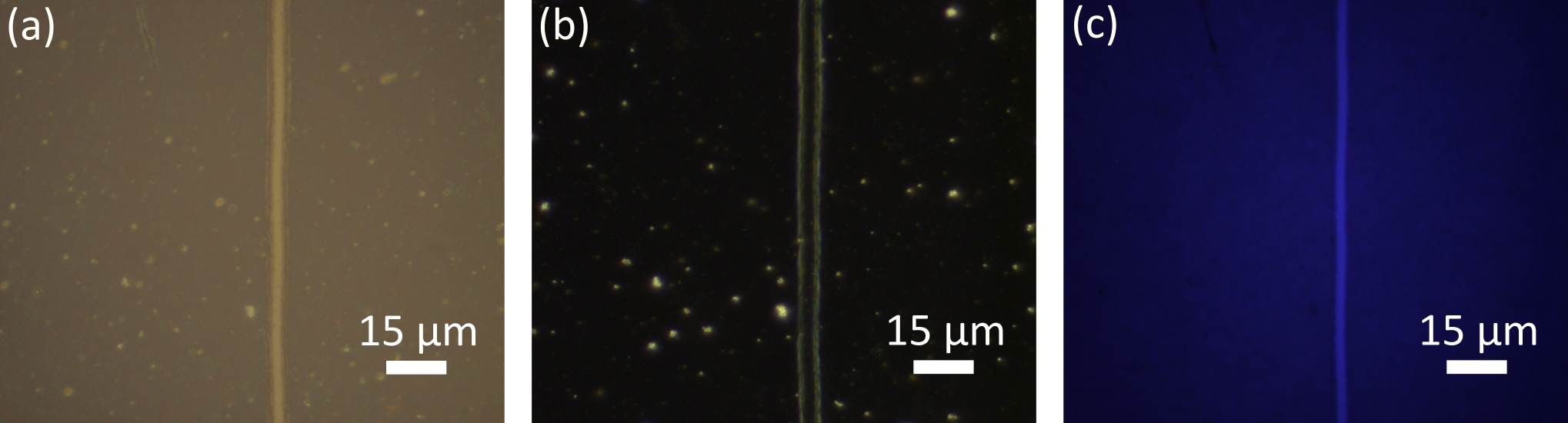}
	\caption{\textcolor{black}{\textbf{SWCNT electrodes.} Bright-field \textbf{(a)}, dark-field \textbf{(b)} and fluorescent dark-field image \textbf{(c)} of the SWCNT electrodes after laser ablation}}
	\label{fig:S3}
\end{figure*}

\begin{figure*}
	\centering
	\includegraphics[scale = 0.75]{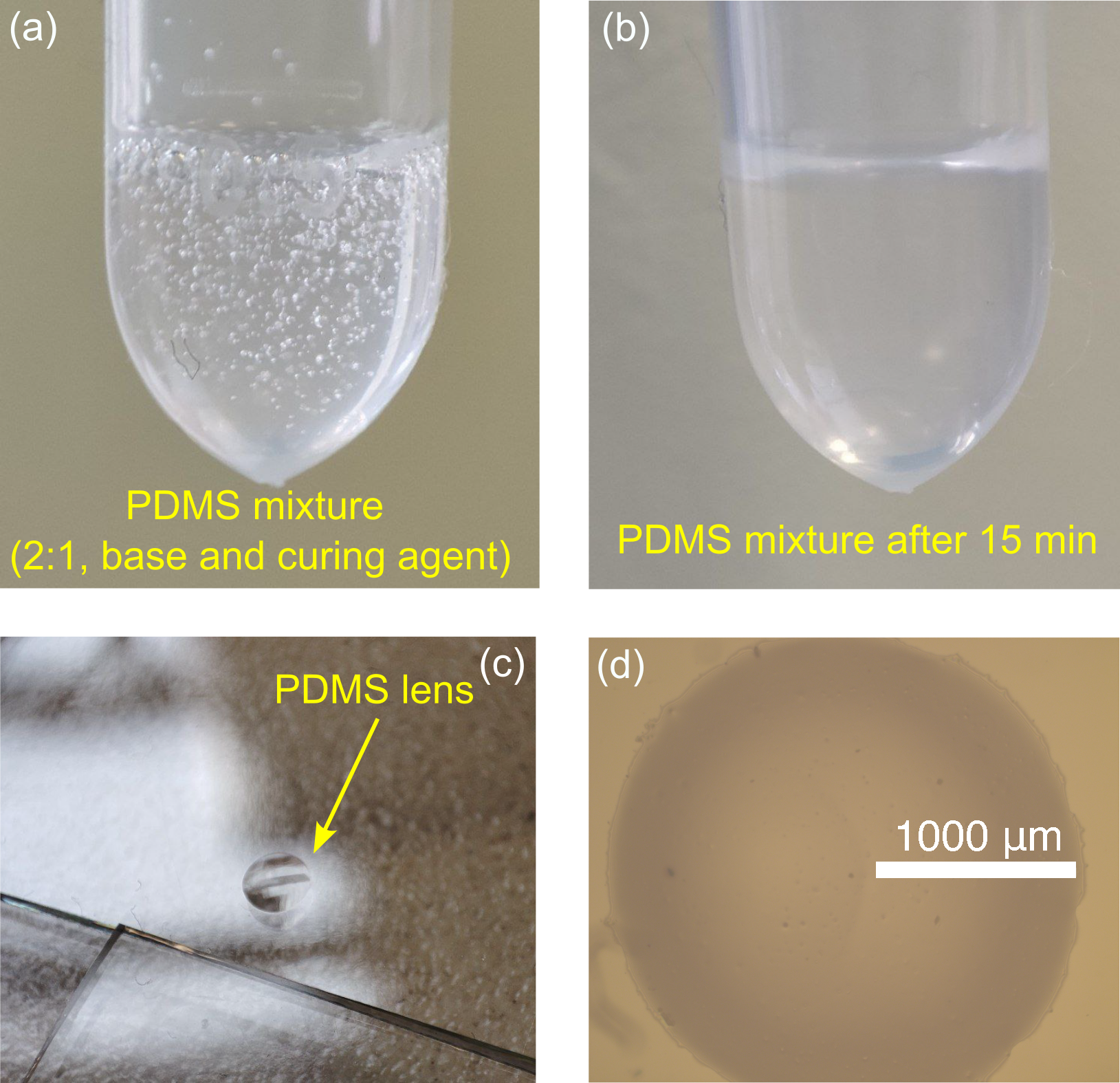}
	\caption{\textcolor{black}{\textbf{Process of making a PDMS lens for the dry transfer process.} (a) Photo of mixture of the base viscous polymer component and a curing agent (2:1 mass ratio) after stirring, (b) photo of mixture after 15 min degassing, (c) photo and bright-field image (d) of PDMS drop on glass susbtrate}}
	\label{fig:S3}
\end{figure*}

\begin{figure*}
	\centering
	\includegraphics[scale = 0.85]{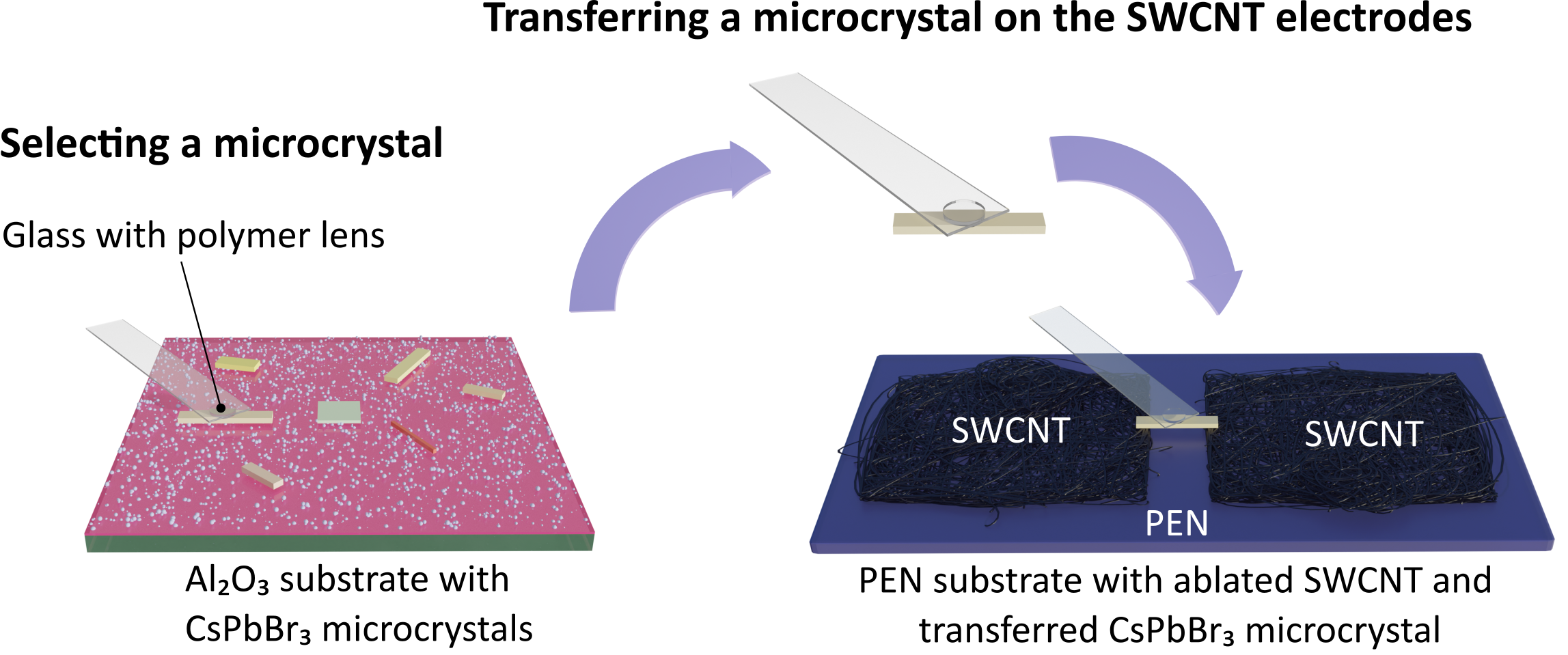}
	\caption{\textcolor{black}{\textbf{Schematic illustration of the transfer process. \textbf{}}}}
	\label{fig:S3}
\end{figure*}

\begin{figure*}
	\centering
	\includegraphics[scale = 0.95]{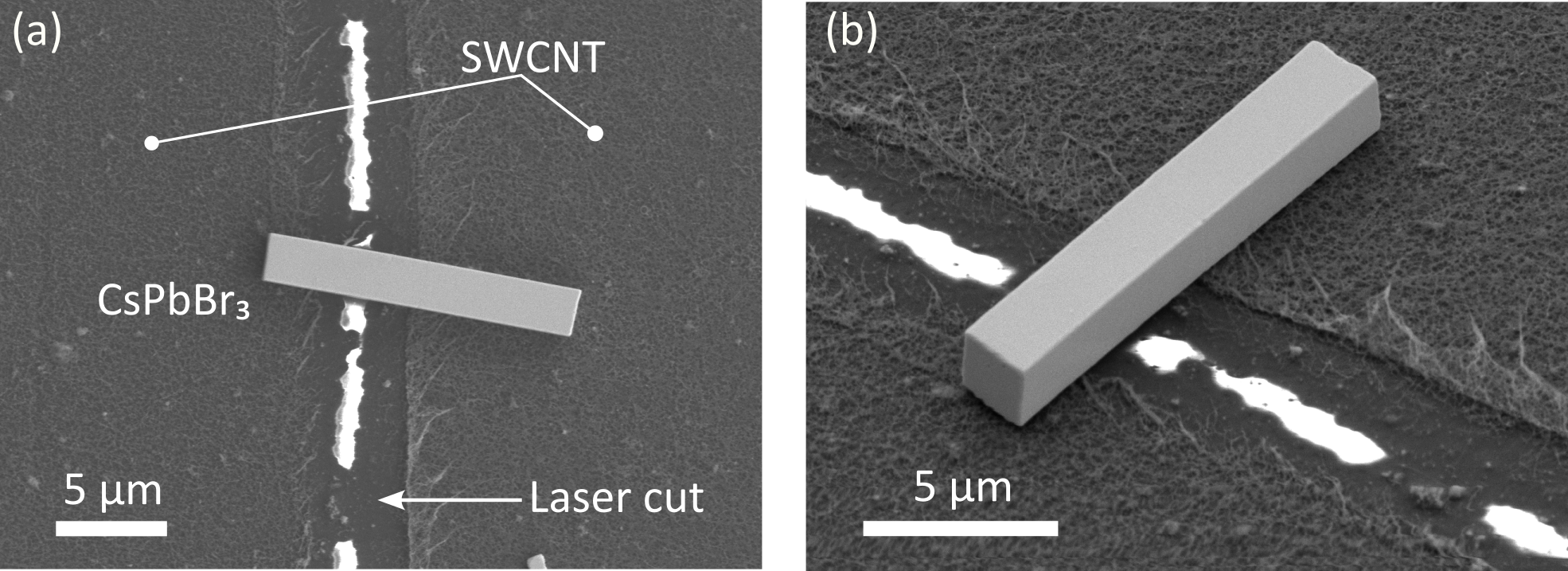}
	\caption{\textcolor{black}{\textbf{Image of the actual device}. \textbf{(a)} Top-view and tilted-angle (b) SEM-image of CsPbBr$_3$ bridging SWCNT electrodes. Image from (b) is magnified Figure 2l from the main text. White lines in the cut area appeared due to the accumulation of charge on the polymer substrate.}}
	\label{fig:S3}
\end{figure*}

\begin{figure*}
    \centering
    \includegraphics[scale = 0.8]{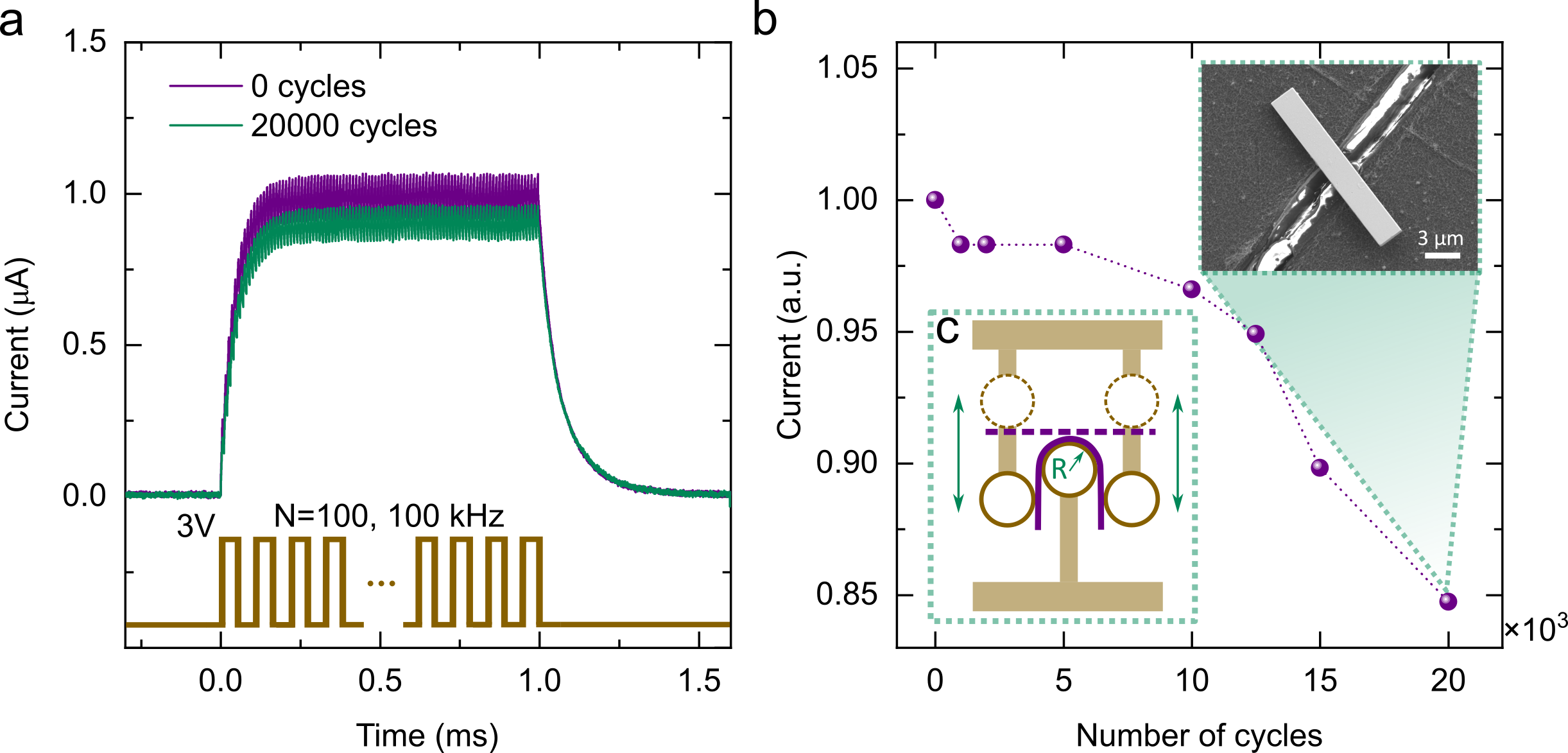}
    \caption{\textbf{Contact stability under optoelectronic stimuli combined with mechanical bending for 20000 cycles.} \textbf{(a)} Optoelectronic stimuli consist of 100 5$\mu$s pulses at 3V and the repetition frequency of 100 kHz and 30 mW/cm$^2$ light illumination at the wavelength of 532 nm. \textbf{(b)} The normalized response of the device to this optoelectrical stimuli under mechanical bending of a tunable number of bending cycles. The device response is measured after each 5000 bending cycles. SEM image (right inset) of the device after 20000 bending cycles shows no visible changes in the morphology of the crystal. \textcolor{black}{White lines in the cut area appeared due to the accumulation of charge on the polymer substrate. \textbf{(c)} Schematic front-view illustration of cylindrical-based bending machine. Curvature radius is equal to R = 2 mm. Arrows indicate direction of movement of cylinders.}}
    \label{fig:s2}
\end{figure*} 

\begin{figure*}
    \centering
    \includegraphics[scale = 1]{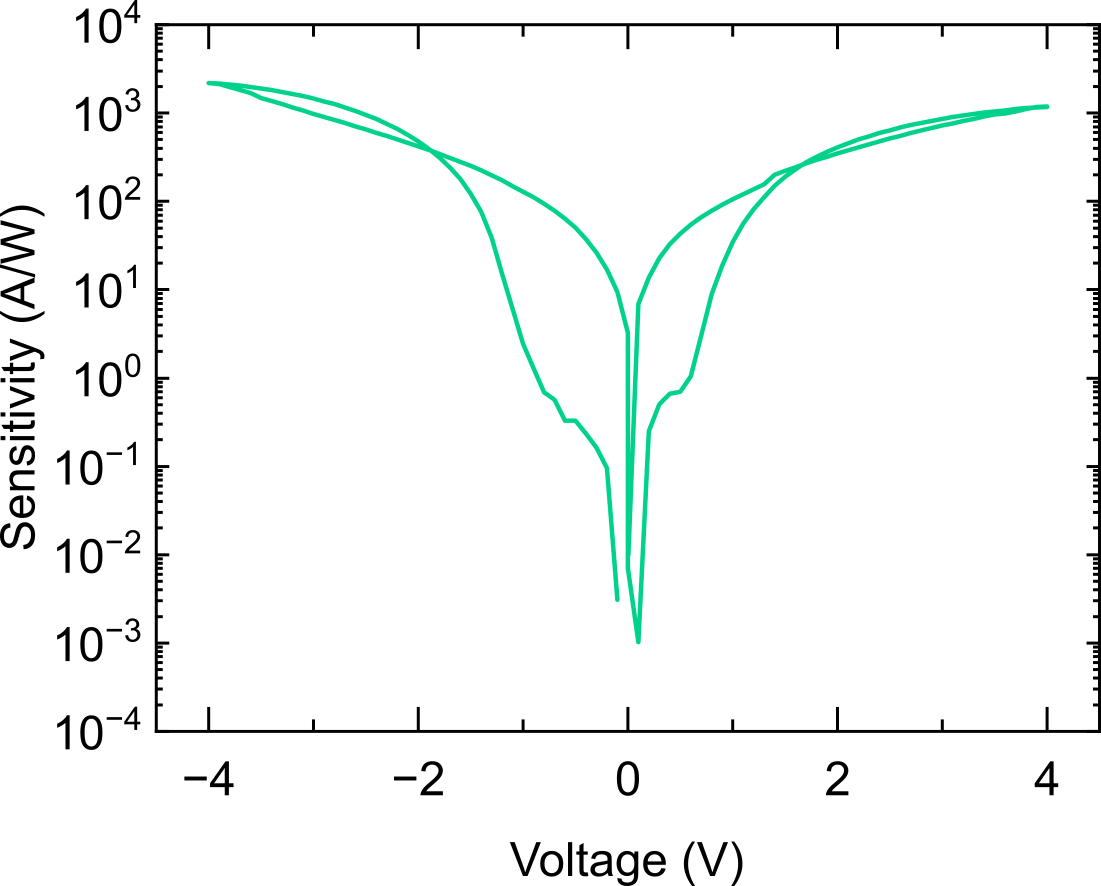}
    \caption{\textbf{Voltage-dependent sensitivity of the photodetecting microcrystal.}}
    \label{fig:s3}
\end{figure*}

\begin{figure*}
    \centering
    \includegraphics[scale = 1]{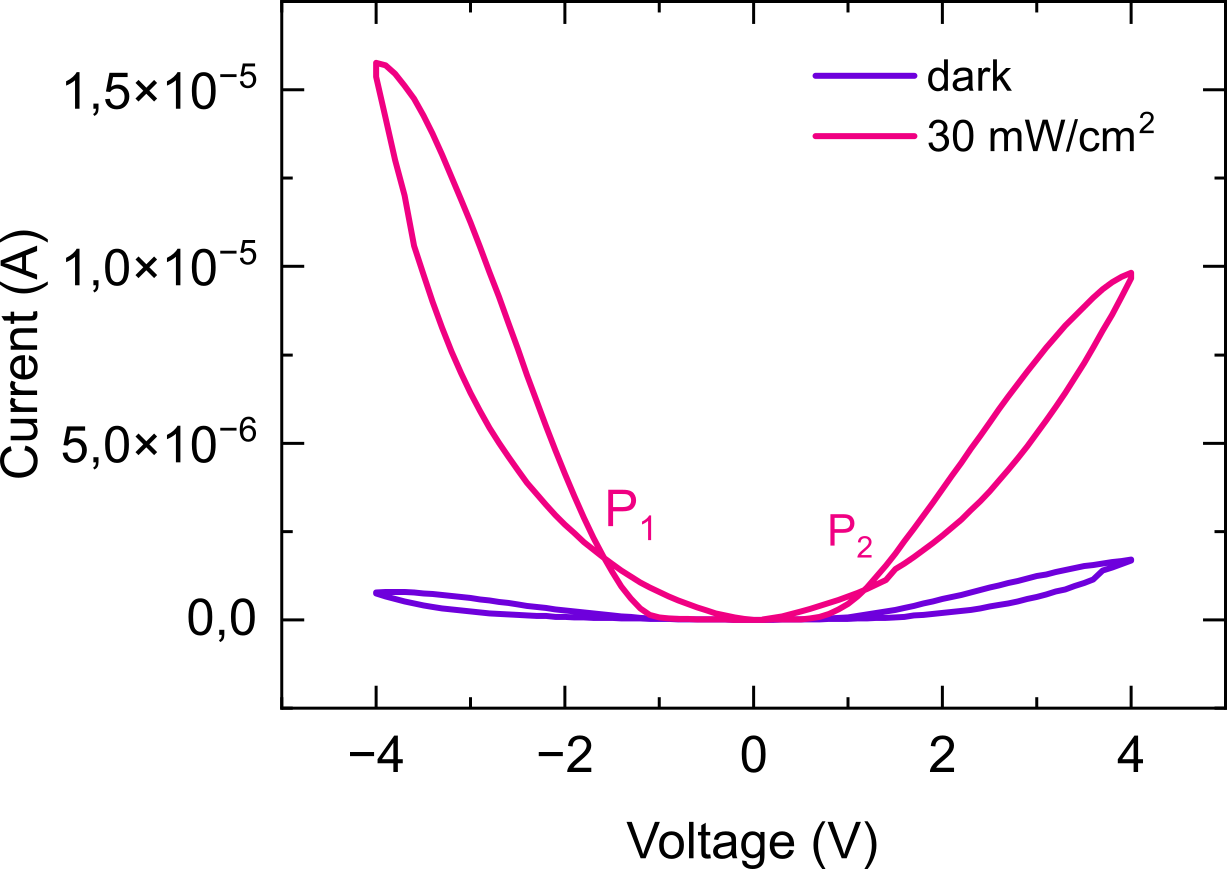}
    \caption{\textbf{Current-voltage in linear scale.}}
    \label{fig:s3}
\end{figure*}

\begin{figure*}
    \centering
    \includegraphics[scale = 0.8]{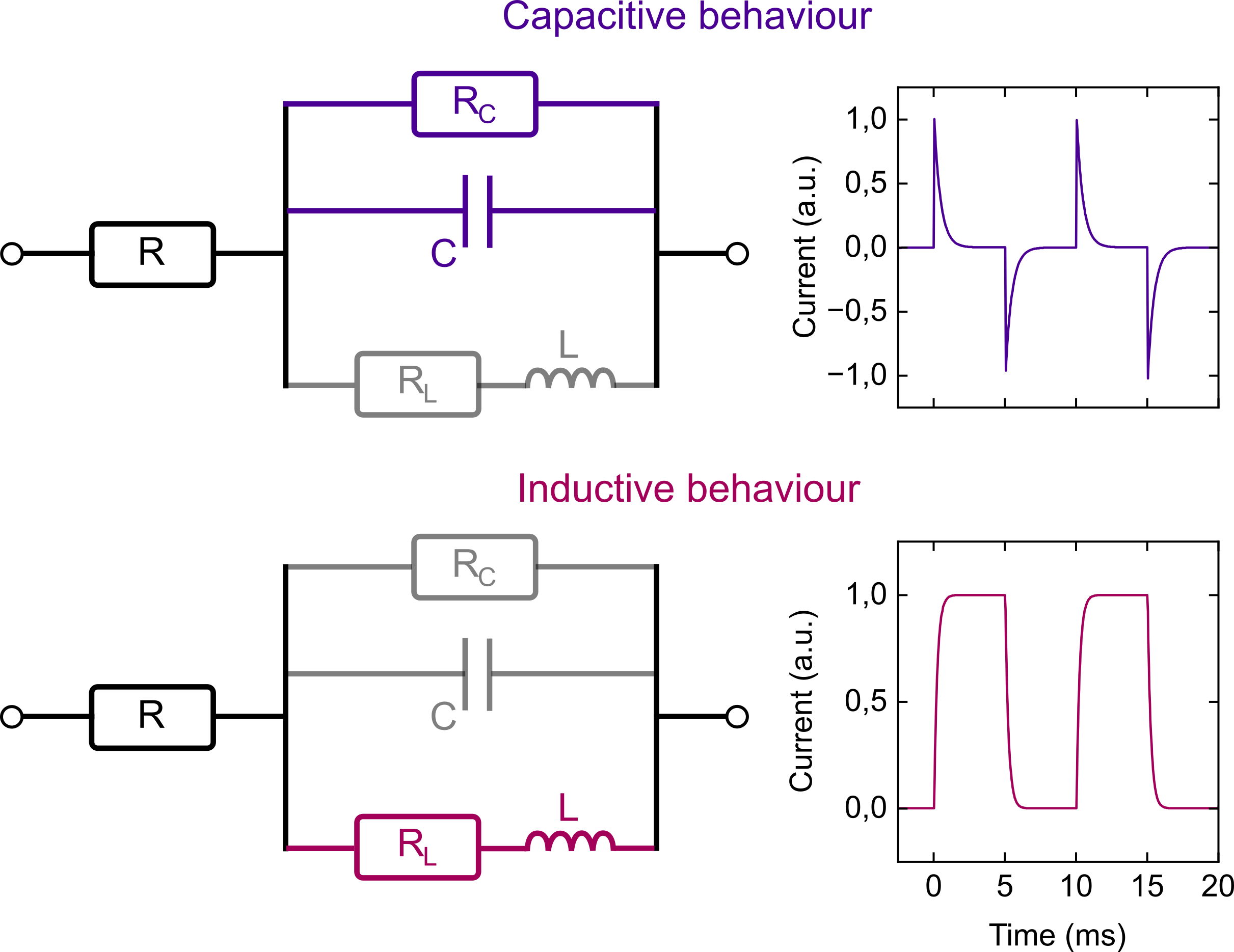}
    \caption{\textbf{Equivalent circuit model with capacitive and inductive branches and corresponding transient responses.}}
    \label{fig:s9}
\end{figure*}

\begin{figure*}
    \centering
    \includegraphics[scale = 0.7]{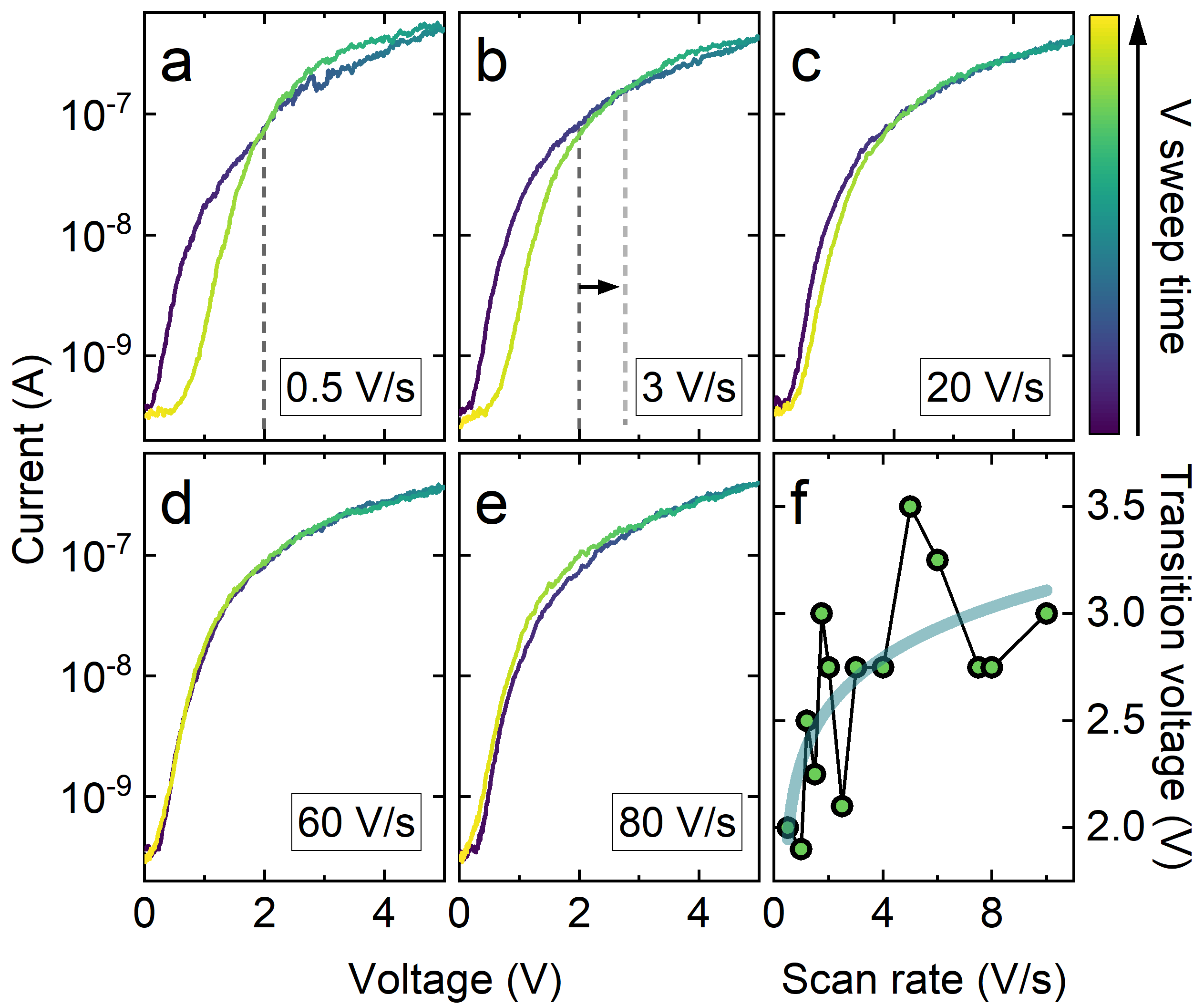}
    \caption{\textcolor{black}{\textbf{Scan-rate analysis of the I-V curve.} \textbf{(a-e)} I-V curves at different scan-rates. \textbf{f} Dependance of the transition voltage (P$_2$ point from Figure 2l of the main text and Figure S12) on the scan rate}}
    \label{fig:s3}
\end{figure*}



\begin{figure*}
    \centering
    \includegraphics[scale = 1]{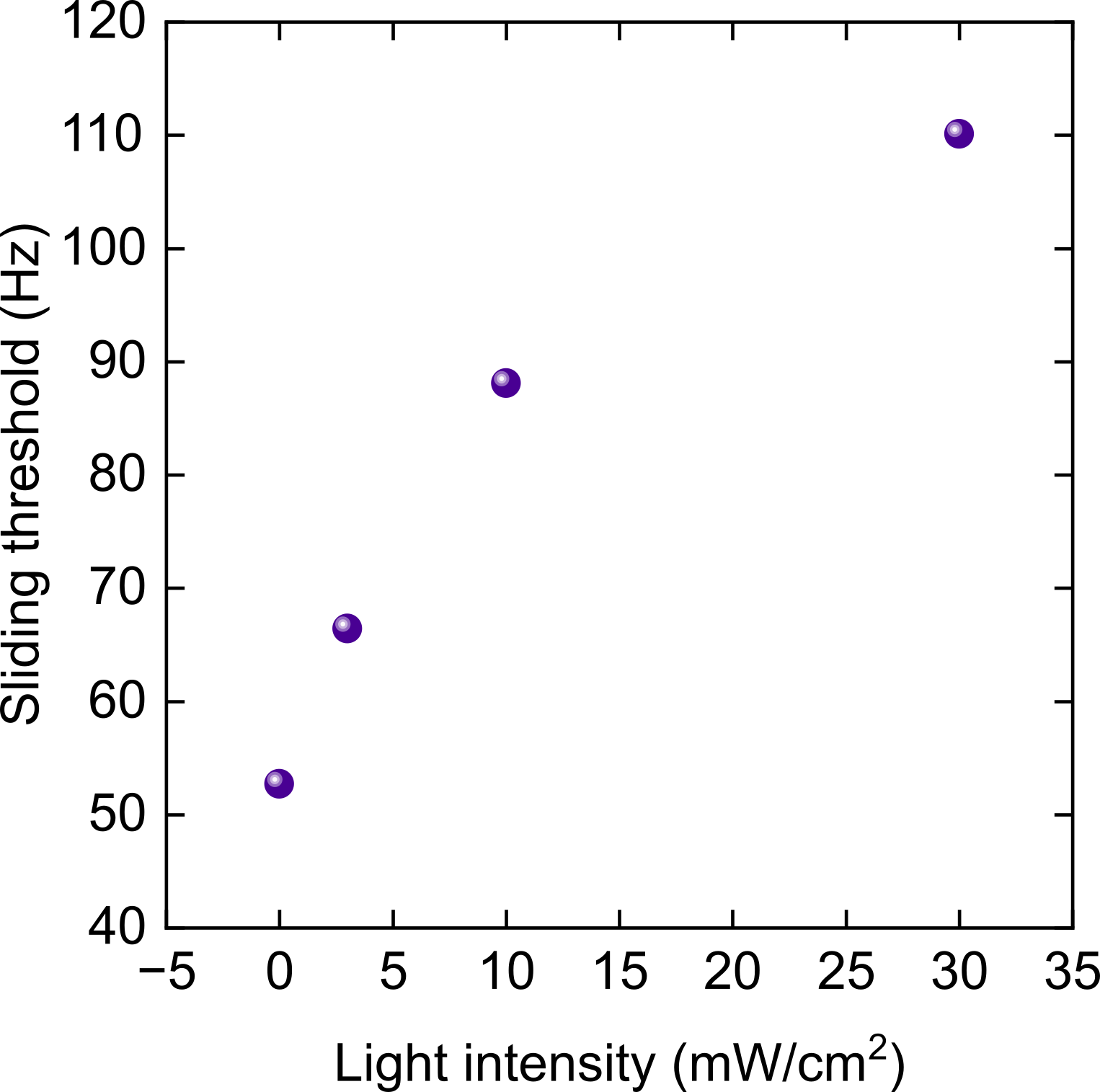}
    \caption{\textbf{The illumination intensity-dependent sliding threshold obtained from BCM experiment}}
    \label{fig:enter-label}
\end{figure*}

\begin{figure*}
    \centering
    \includegraphics[scale = 1]{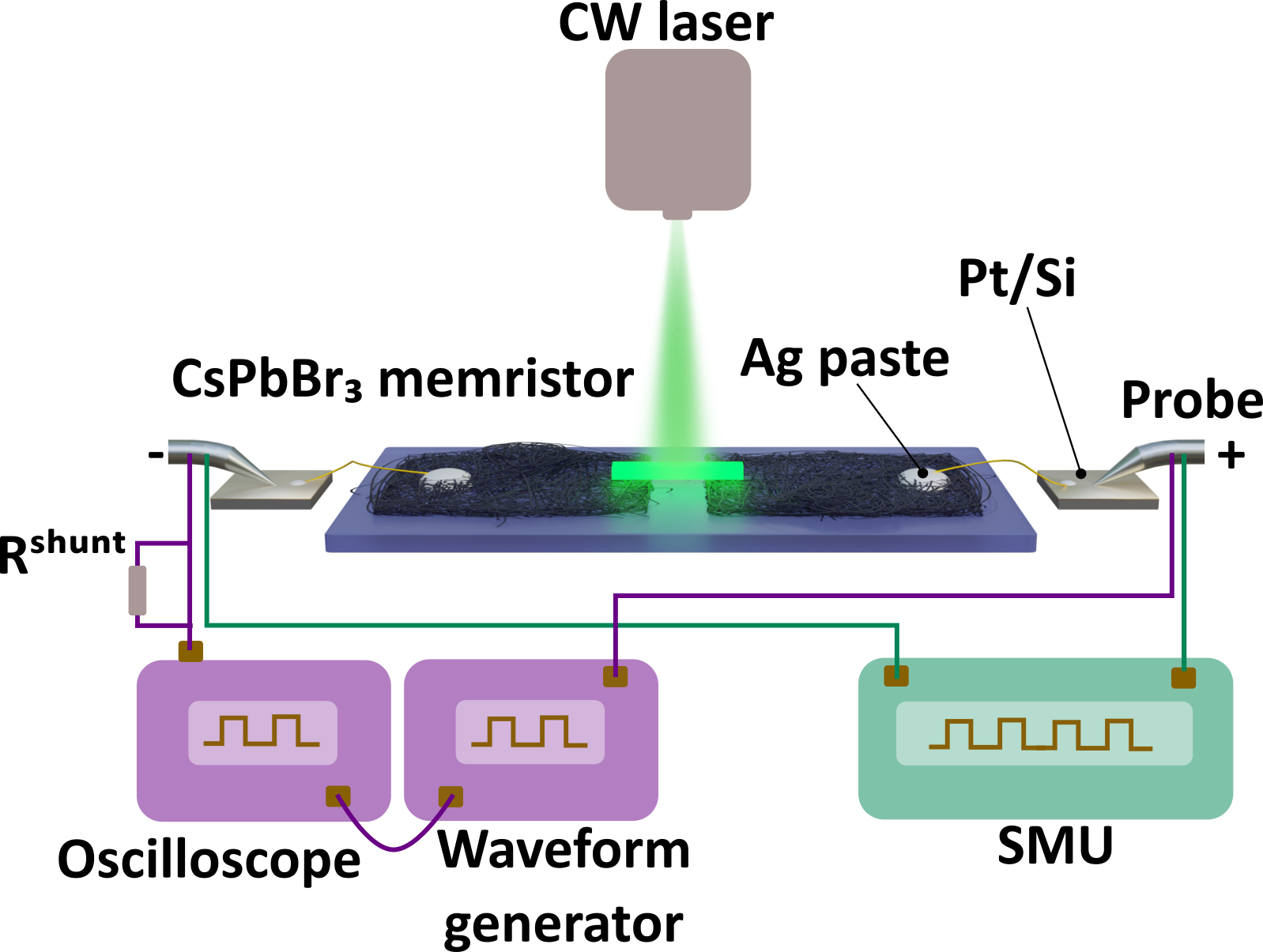}
    \caption{\textcolor{black}{\textbf{Schematic setup for the optoelectrical experiments.}}}
    \label{fig:enter-label}
\end{figure*}

\newpage
